\documentclass[aps, prc, twocolumn, amsmath, amssymb, superscriptaddress]{revtex4}
\usepackage{mathptmx}
\usepackage{CJK}
\usepackage{graphicx}
\usepackage{bm}
\usepackage{dcolumn}
\usepackage{multirow}
\usepackage{xcolor}
\usepackage[
breaklinks,
pdfstartview=FitH,
CJKbookmarks=true,
bookmarksnumbered=true,
bookmarksopen=true,
colorlinks=true,
linkcolor=blue,
anchorcolor=blue,
citecolor=blue
]{hyperref}

\DeclareMathAlphabet{\mathcal}{OMS}{cmsy}{m}{n}

\def\beq{\begin{equation}} \def\eeq{\end{equation}}
\def\bal#1\eal{\begin{align}#1\end{align}}
\def\bse#1\ese{\begin{subequations}#1\end{subequations}}

\def\al{\alpha}
\def\be{\beta}

\def\la{\Lambda}

\def\fm3{\,\text{fm}^{-3}}

\def\mfm5{\;\text{MeV}\,\text{fm}^5}

\def\c12{$^{12}$C}

\def\pnl{\Phi^{(N\la)}\!}

\begin{document}

\begin{CJK*}{UTF8}{gbsn}
\title{Large quadrupole moment in $^{20}$Ne: 
Localization and tensor-force effects}

\author{Huai-Tong~Xue~(薛怀通)}
\thanks{These authors contributed equally to this work.}
\affiliation{College of Physics and Electronic
Engineering, Nanyang Normal University, Nanyang 473061, China}

\author{Suo~Qiu~(裘索)} 
\thanks{These authors contributed equally to this work.}
\affiliation{Department of Physics, 
East China Normal University, Shanghai 200241, China}

\author{C.-F.~Chen~(陈超锋)} %
\thanks{These authors contributed equally to this work.}
\affiliation{School of Physics Science and Engineering, 
Tongji University, Shanghai 200092, China}

\author{Q.~B.~Chen~(陈启博)}
\affiliation{Department of Physics, 
East China Normal University, Shanghai 200241, China}

\author{Xian-Rong~Zhou~(周先荣)}\email[Contact author:~]{xrzhou@phy.ecnu.edu.cn}
\affiliation{Department of Physics, 
East China Normal University, Shanghai 200241, China}

\author{Zhongzhou~Ren~(任中洲)} %
\affiliation{School of Physics Science and Engineering, 
Tongji University, Shanghai 200092, China}

\date{\today}

\begin{abstract}

Using the Beyond-Skyrme-Hartree-Fock approach with 
various Skyrme-type $NN$ effective interactions, 
we explore the large spectroscopic quadrupole moment
$Q_s$ of the first excited $2_{1}^{+}$ state in $^{20}$Ne. 
Our calculated $Q_s(2_{1}^{+}) \approx -0.20~e\textrm{b}$ 
aligns closely with the experimental value of $-0.22(2)~e\textrm{b}$. 
Consistent with other theoretical methods, our results 
confirm the presence of an $\alpha$ cluster structure
in this state. We also investigate the robustness of
the $\alpha$ cluster against tensor-force, demonstrating 
that incorporating tensor components significantly enhances 
$Q_s(2_{1}^{+})$ and strengthens the cluster structure 
along the symmetrical axis.

\end{abstract}

\maketitle

\section{Introduction}
\label{s:intro}
The static (spectroscopic) electric quadrupole moments of 
nuclei serve as critical indicators of deformation properties
within nuclear systems~\cite{Carchidi1986prc}. Their 
sign reversals and magnitude fluctuations across different
neutron number ($N$) and proton number ($Z$) directly reveal 
the collective behavior and multi-nucleon configurations 
inherent in nuclear quantum states~\cite{Carchidi1986prc}. 
This phenomenon highlights how intricate interactions 
between nucleons govern the spatial charge distribution 
patterns observed in excited or ground-state nuclei. 
It is a fundamental challenge for a microscopic theory, 
such as the beyond-mean-field model, to account 
for the nuclear shape features reflected by these data.

The measurement of nuclear quadrupole moments has always 
been more challenging than that of magnetic moments~\cite{Neyens2003RPP}. 
While the first quadrupole moments for stable nuclei 
were reported in the 1960s, large-scale studies began 
in the late 1970s, relying on muonic X-ray hyperfine 
structure analysis and atomic beam magnetic resonance. 
Measurements for long-lived radioactive nuclei also 
emerged during this period. Several key 
compilations~\cite{Lederer1978tableof, Raghavan1989ADNDT,
Stone2005ADNDT} catalog over 40 experimental 
techniques that account for the wide variety of nuclear
properties, such as lifetimes, spin values, chemical 
characteristics, production mechanisms, and decay modes. 
In addition, dedicated studies have focused on specific 
nuclear regions~\cite{Loebner1970NDSA, Christy1973NDSA,
Spear1981PR, Neyens2003RPP}. 

More recently, interest in nuclear shape and deformation 
has extended into the field of relativistic heavy-ion 
collision (HIC). In particular, light-ion collision (LIC) 
has emerged as a novel research avenue at modern HIC 
facilities, aimed at probing both the hydrodynamic 
behavior of the quark-gluon plasma and the geometric 
features of light nuclei~\cite{mehl2024prc}. In such 
collision, shape can manifest through observables 
related to the initial overlap region and collision 
centrality~\cite{Aaij2022JoI}. LICs involving symmetric
light nuclei (e.g., $^{16}$O+$^{16}$O and $^{20}$Ne+$^{20}$Ne)
offer advantages over conventional Pb+Pb or Au+Au systems, 
due to reduced sensitivity to spatial fluctuations 
in protons and more well-defined centrality~\cite{mehl2024prc}. 
In particular, $^{20}$Ne has attracted significant attentions, 
as its initial cluster-like geometry, often described as 
a ``bowling pin" shape, may be inferred from deviations 
in the incoherent cross section compared to spherical 
assumptions~\cite{Ebran2012nature, Mantysaari2023PRL}. 
Understanding such initial-state deformations is
crucial for interpreting future LIC data and 
exploring clustering phenomena in light nuclei.

Earlier Coulomb excitation measurements of 
$^{20}$Ne~\cite{Nakai1970NPA, Schwalm1972NPA, Olsen1974NPA} 
yielded three values of the electric quadrupole 
moment $Q_s(2_1^+)$. The weighted average 
of these results, $Q_s(2_1^+) = -0.23(3)~e\textrm{b}$, 
was accepted by the National Nuclear Data Center (NNDC)~\cite{NuDat}. 
This value represents one of the largest prolate 
deformations in the nuclear chart, corresponding 
to a quadrupole parameter of $\beta_2 = 0.96(11)$~\cite{bm}.

The large $Q_s(2_1^+)$ in $^{20}$Ne has been a long-standing 
challenge in nuclear structure theory~\cite{thomson1913bakerian}.
This state is often considered to exhibit one of 
the most extreme quadrupole shapes in the nuclear chart. 
Despite the presence of only two protons and two neutrons above 
the shell closure at magic number $8$, which suggests a favorable 
prolate deformation for the $sd$ shell due to the relatively 
sharper energy decrease of the $[220 1/2]$ single-particle 
Nilsson orbit as a function of deformation~\cite{bm}. 
However, nuclear structure models still struggle to reproduce 
the large $Q_s$. For example, predictions from the shell model~\cite{Spear1981PR,
Carchidi1986prc}, the rotational model of Bohr and Mottelson~\cite{bm}, 
beyond-mean-field models (including the Gogny energy density 
functional~\cite{20Negogny}, relativistic energy density 
functional~\cite{EFzhou2016PLB}, and relativistic Hartree 
Bogoliubov approach~\cite{marevic2018prc}), the resonating 
group method~\cite{matsuse1975PTP}, multi configurational 
models~\cite{lebloas2014prc}, and $ab~initio$ calculations 
using the in-medium similarity renormalization group 
(VS-IMSRG)~\cite{Hergert2016PR, Stroberg2019AR, Bogner2014PRL} 
tend to underestimate $Q_s$.

Recently, the $Q_s(2_1^+)$ of $^{20}$Ne was reexamined through 
both experimental and theoretical approaches~\cite{mehl2024prc}. 
A precise value of $-0.22(2)~e\textrm{b}$ was determined using 
RECE at backward angles, consistent with earlier 
measurements~\cite{Schwalm1972NPA}. Several theoretical 
models were employed to reproduce the data, including 
the shell model with WBP interaction, the $ab~initio$ 
shell model with chiral effective interactions, and 
the multi-reference relativistic energy density 
functional (MR-EDF) model. The relativistic
MR-EDF model, which accounts for $\alpha$ clustering, 
provides a superior prediction ($\approx -0.15~e\textrm{b}$) 
compared to other models that do not incorporate 
$\alpha$ clustering.

In parallel, non-relativistic beyond-mean-field approaches 
based on the Skyrme density functional (Beyond-SHF model) 
have proven to be effective in studying low-lying properties 
of normal nuclei~\cite{Bonche1990npa, Bender2006prc, Dobaczewski09, 
schunck2012CPC} and hypernuclei~\cite{Cui2017prc, Li2018prc, 
Xue2024prc2, Xue2024prc, C.F.Chen_cpc_2022, J.Guo_prc_2022, Y.Liu2023prc}. 
It is well known that non-relativistic mean-field 
models generally predict shallower nucleon potentials compared to relativistic 
mean-field models, making it less likely for density distributions
to exhibit clustering explicitly~\cite{Ebran2012nature}. 
However, recent studies suggest that structures resembling 
$\alpha$ clusters can be observed in the ground state of $^{20}$Ne 
using localization functions~\cite{Reinhard2011prc} that 
has been used in several studies of nuclear structure 
and reaction~\cite{C.L.Zhang2016prc, Schuetrumpf2017prc, 
Mercier2021prc, Li2024prc}, opening the way of
studying $\alpha$ cluster or localization in the 
Skyrme model.

Furthermore, one notes that the non-central interactions 
are likely critical in enhancing nuclear deformation~\cite{Sagawa2014PPNP}. 
To improve the agreement between theoretical predictions 
and experimental data for the quadrupole moment,
we will investigate the impact of incorporating 
such interactions. tensor-force, a prominent type 
of non-central interaction, significantly influence 
nuclear structure and reactions~\cite{Sagawa2014PPNP, 
Godbey2019prc}. Early implementations of tensor 
forces added a tensor term to the nucleon-nucleon 
interaction without refitting other parameters, 
adjusting only the tensor strength to match experimental 
data. Examples include the SLy5+T~\cite{Colo2007PLB} 
and the SIII+T~\cite{Brink2007PRC} 
parameterizations. More recent approaches employ 
a fully self-consistent strategy, dynamically 
optimizing all parameters within a defined tensor 
strength space, as seen in the 36 independent 
parametrizations of the TIJ family~\cite{TIJ}.
Another notable example is the SAMi+T~\cite{SAMiT} 
interaction, which incorporates tensor-forces derived 
from $ab~initio$ relativistic Brueckner-Hartree-Fock 
(RBHF) calculations of neutron-proton drops.

Therefore, in this work, we will apply the Beyond-SHF model to investigate the 
correlation between the pronounced quadrupole deformation and the underlying 
$\alpha$-cluster structure in the $2_1^+$ state of $^{20}$Ne, 
with a particular focus on the role of tensor forces. 
We first compare theoretical predictions with experimental data to assess the model’s 
capability in describing low-lying excited states in the sd-shell region. Based on this comparison, 
we then conduct a detailed analysis of how tensor interactions affect the low-lying spectra 
and the spectroscopic quadrupole moment $Q_s(2_1^+)$.

\section{Theoretical framework}
\label{s:theo}

\subsection{Localization function in beyond-mean-field}

In the beyond SHF model, nuclear states are described by a superposition of angular-momentum-projected mean-field wavefunctions.
\begin{equation}\label{e:psi}
\left|\Psi_\alpha^{J M}\right\rangle=\sum_\beta F_\alpha^J(\beta) \hat{P}_{M K}^J\left|\Phi(\beta)\right\rangle
\end{equation}
where $F_\alpha^J(\beta)$ is a weight function, and $\hat{P}_{M K}^J$ is the angular momentum projection~(AMP) operator, with $K$ representing the projection of angular momentum $J^\pi$ onto the intrinsic $z$ axis.

To obtain the eigenstate $\left|\Psi_\al^{JM}\right\rangle$,
each weight $F_\al^J(\be)$ in Eq.~(\ref{e:psi})
is determined by the Hill-Wheeler-Griffin (HWG) equation~\cite{Peter1980},
\beq
\sum_{\be'}
\Big[ H_{KK}'^J(\be,\be') - E_\al^J N_{KK}^J(\be,\be') \Big]
F_\al^J(\be') = 0 \:,
\label{e:hwg}
\eeq
in which the Hamiltonian and norm elements are given by
\bal
H_{KK'}'^J(\be,\be') &=
\big\langle \pnl(\be') \left| \hat{H}' \hat{P}_{KK'}^J \right|
\pnl(\be) \big\rangle \:,
\\
N_{KK'}^J(\be,\be') &=
\big\langle \pnl(\be') \left| \hat{P}_{KK'}^J \right|
\pnl(\be) \big\rangle \:.
\label{e:nkk}
\eal
The Hamiltonian $\hat{H}'$ is defined as
\beq
\hat{H}' = \hat{H}
- \lambda_p (\hat{N}_p-Z) - \lambda_n (\hat{N}_n-N) \:,
\eeq
where the Hamiltonian $\hat{H}$ is determined by the nuclear EDF,
and the last two terms account for the fact
that the projected wave function
does not provide the correct number of particles on average
\cite{Bonche1990npa}.   

With the wave functions, the static electric quadrupole
moments $Q_s(2_1^+)$ read,
\begin{equation}\label{eq:qs}
Q_s(2_1^+)=\langle 2_1^+|| \hat{Q}_2 || 2_1^+ \rangle,
\end{equation}
where the reduced matrix elements of quadrupole moment are,
\begin{align}
	\langle J_{\alpha^{\prime}}^{\prime+}|| 
	\hat{Q}_2||J_\alpha^{+}\rangle
	& = \sqrt{2 J^{\prime}+1} 
	\sum_{M \mu \beta \beta^{\prime}} 
	F_{\alpha^{\prime}}^{J^{\prime}}(\beta^{\prime})^* 
	F_\alpha^J(\beta) C_{J M 2 \mu}^{J^{\prime} K^{\prime}} \notag \\
	& \times \langle\Phi(\beta^{\prime})| 
	\hat{Q}_{2 \mu} \hat{P}_{M K}^J|\Phi(\beta)\rangle,
\end{align}
with $C_{J M 2 \mu}^{J^{\prime} K^{\prime}}$ 
denoting the Clebsh-Gordon coefficients. 
The electric quadrupole transition operator~\cite{Bonche1990npa}
is
\begin{align}
	\hat{Q}_{2 \mu}=\sum_k e_k r_k^2 Y_{2 \mu}(\theta_k, \varphi_k),
\end{align}
where $e_k$ is the charge of the $k~\textrm{th}$ nucleon 
and ($r_k$, $\theta_k$, $\varphi_k$) is its position 
relative to the center of mass of the nucleus. 
Bare charges are used in this calculation
(i.e., $e_p=e$ and $e_n=0)$.

To describe the probability of finding two like-spin particles in the vicinity of each other, 
a normalized localization measure that depends on the local density $\rho(\mathbf{r,\sigma})$, 
gradient of local density $\nabla \rho(\mathbf{r,\sigma}) $, 
kinetic energy density$\tau(\mathbf{r,\sigma})$, and current density $\mathbf{j(\mathbf{r,\sigma})}$ has been adopted in the nuclear systems~\cite{Reinhard2011prc}.

Because the reference states $|\Phi(\beta)\rangle$ mixed by generator coordinate method (GCM) are various static configurations 
obtained from the mean field calculation and AMP. So, the localization function of a state 
with definite angular momentum $J$ can be expressed as
\begin{equation}
C_{q \sigma}^J(\mathbf{r})=\left[1+\left(\frac{\tau_{q \sigma}^J(\mathbf{r}) \rho_{q \sigma}^J(\mathbf{r})-\frac{1}{4}\left|\nabla \rho_{q \sigma}^J(\mathbf{r})\right|^2-\left|\mathbf{j}_{q \sigma}^J(\mathbf{r})\right|^2}{\rho_{q \sigma}^J(\mathbf{r}) \tau_{q \sigma}^{\mathrm{TF}, J}(\mathbf{r})}\right)^2\right]^{-1}
\end{equation}

with
\begin{equation}
\begin{aligned}
\rho_{q \sigma}^{J} (\mathbf{r})&=\sum_{\beta^{\prime} \beta} F_{\alpha}^{J}(\beta^{\prime})^* 
F_\alpha^J(\beta) 
\left\langle\Phi\left(\beta^{\prime}\right)\right| \hat{\rho}(\mathbf{r}) \hat{P}_{KK}^{J}\left|\Phi\left(\beta\right)\right\rangle,\\
\tau_{q \sigma}^{J} (\mathbf{r})&=\sum_{\beta^{\prime} \beta} F_{\alpha}^{J}(\beta^{\prime})^* 
F_\alpha^J(\beta)
\left\langle\Phi\left(\beta^{\prime}\right)\right| \hat{\tau}(\mathbf{r}) \hat{P}_{KK}^{J}\left|\Phi\left(\beta\right)\right\rangle,\\
\mathbf{j}_{q \sigma}^{J} (\mathbf{r})&=\sum_{\beta^{\prime} \beta} F_{\alpha}^{J}(\beta^{\prime})^* 
F_\alpha^J(\beta)
\left\langle\Phi\left(\beta^{\prime}\right)\right| \hat{j}(\mathbf{r}) \hat{P}_{KK}^{J}\left|\Phi\left(\beta\right)\right\rangle,\\
\tau_{q \sigma}^{\mathrm{TF}}&=\frac{3}{5}\left(6 \pi^2\right)^{2 / 3} {\rho_{\alpha}^{J} }^{5 / 3}.
\end{aligned}
\end{equation}
The subscript label $q$ denotes neutrons (protons).
$C_{q \sigma}^J(\mathbf{r})$ can be interpreted as a quantitative measure representing the extent 
of localization of quantum  states within atomic nuclei.

In the present work, we focus on the even-even nucleus $^{20}$Ne, 
for which both time-reversal symmetry and signature symmetry are preserved in the calculations. 
As a consequence, the physical quantities for a pair of time-reversed 
single-particle states (i.e., with opposite spin orientations but otherwise identical quantum numbers) are identical. 
Therefore, the localization function for one spin projection (e.g., spin-up) is sufficient to represent that of the opposite spin. 
Furthermore, since the number of neutrons and protons is equal in $^{20}$Ne, the localization functions for neutrons and protons 
are expected to be similar. In the following, we thus present and discuss the results for spin-up neutrons only.
In addition, the current density $\mathbf{j(\mathbf{r})}$ vanishes in the static case~\cite{Reinhard2011prc}.
Even if we do AMP and configuration mixing, 
the entire system is still ``static", so the $\mathbf{j(\mathbf{r})}$ term can still be ignored in the time-reversal symmetry  even-even nuclei.

\subsection{A brief introduction to tensor force}
The tensor term, also known as the $\mathcal{J}^2$ term, originates from both the zero-range central 
and tensor forces contribution to the energy functional is given by. 
By considering these two types of forces, $\varepsilon_{\text {Tensor }}$ can be written as

\begin{equation}
	\varepsilon_{\text {Tensor }}(r)=\frac{1}{2} \alpha\left(\mathbf{J}_q^2+\mathbf{J}_{q^{\prime}}^2\right)+\beta \mathbf{J}_q \mathbf{J}_{q^{\prime}},
\end{equation}
where $\alpha$ and $\beta$ are the like-particle and proton-neutron coupling constants, respectively. The subscript label $q$ denotes neutrons (protons) and $q^{\prime}$ represents protons (neutrons).

The proton-neutron coupling constants $\alpha=\alpha_C+\alpha_T$ and $\beta=\beta_C+\beta_T$ can again be separated into contributions from central and tensor forces,
\begin{equation}\label{eq:al-be}
	\begin{aligned}
	\alpha_C & =\frac{1}{8}\left(t_1-t_2\right)-\frac{1}{8}\left(t_1 x_1+t_2 x_2\right), \\
	\beta_C & =-\frac{1}{8}\left(t_1 x_1+t_2 x_2\right), \\
	\alpha_T & =\frac{5}{4} t_o=\frac{5}{12} U, \\
	\beta_T & =\frac{5}{8}\left(t_e+t_o\right)=\frac{5}{24}(T+U) .
	\end{aligned}
\end{equation}

The values of $\alpha$ and $\beta$ are obtained from various Skyrme energy density functionals, 
which can be grouped into three categories.
(1) Functionals without explicit tensor terms, i.e., these functionals omit the tensor contributions $\alpha_T$ and $\beta_T$, 
but do account for the $\mathbf{J}^{2}$ term arising from the central interaction. 
Examples include SkP~\cite{skp}, SLy5~\cite{Chabanat1998npa}, SkO$^{\prime}$~\cite{sko'}, BSk9~\cite{BSk9}, and SAMi~\cite{SAMi}.
(2) Functionals with perturbative tensor inclusion. In this group, tensor terms are added in a 
perturbative manner without re-adjusting the original parameter sets. Representative cases are SLy5~\cite{Colo2007PLB} 
and SIII+T~\cite{Brink2007PRC}.
(3) Functionals with fully fitted tensor terms, namely tensor contributions are treated on the same level 
as other terms during the fitting process. This category includes Skxta and Skxtb~\cite{Skxtb}, 
the TIJ family (with I, J ranging from 1 to 6)~\cite{TIJ}, SkP-T, SLy4-T, SkO-T~\cite{SLy4T}, and SAMi-T~\cite{SAMiT}.

For the TIJ family, the tensor effect can be shown by the evolution of nuclear structure with $I$ or $J$, 
and the the $\alpha$, $\beta$ is given by
\begin{equation}\label{eq:al-be-IJ}
 \alpha=60(J-2), \quad	\beta=60(I-2) .
\end{equation}

\section{Results and discussion}
\label{s:results}
In the current Beyond-SHF model, only the quadrupole deformation 
parameter $\beta$ of the nuclear core serves as the generator 
coordinate, with intrinsic wave functions maintained as axially 
symmetric. The model space is defined by the range of $\beta$,
which spans from $-1.5$ to $3.0$, and a total of 76 basis
functions are utilized for $^{20}$Ne. In the calculations, several 
Skyrme functionals are employed, including SLy4~\cite{Chabanat1998npa}, 
SIII~\cite{Beiner1975npa}, SKM$^{\prime}$~\cite{Chabanat1998npa},  
SLy5~\cite{Chabanat1998npa}, UDF0~\cite{UDF0}, SAMi~\cite{SAMi}, 
and SGII~\cite{Vangiai1981npa}. 
The pairing force strength is set to $V_0 = -410~\textrm{MeV}\textrm{fm}^3$ 
for both protons and neutrons~\cite{Sagawa2004PRC}, with
a smooth pairing energy cutoff of $5~\textrm{MeV}$ 
around the Fermi level, as in Ref.~\cite{Terasaki96}.

\begin{figure}[ht]
\begin{center}
  \includegraphics[width=0.85\linewidth]{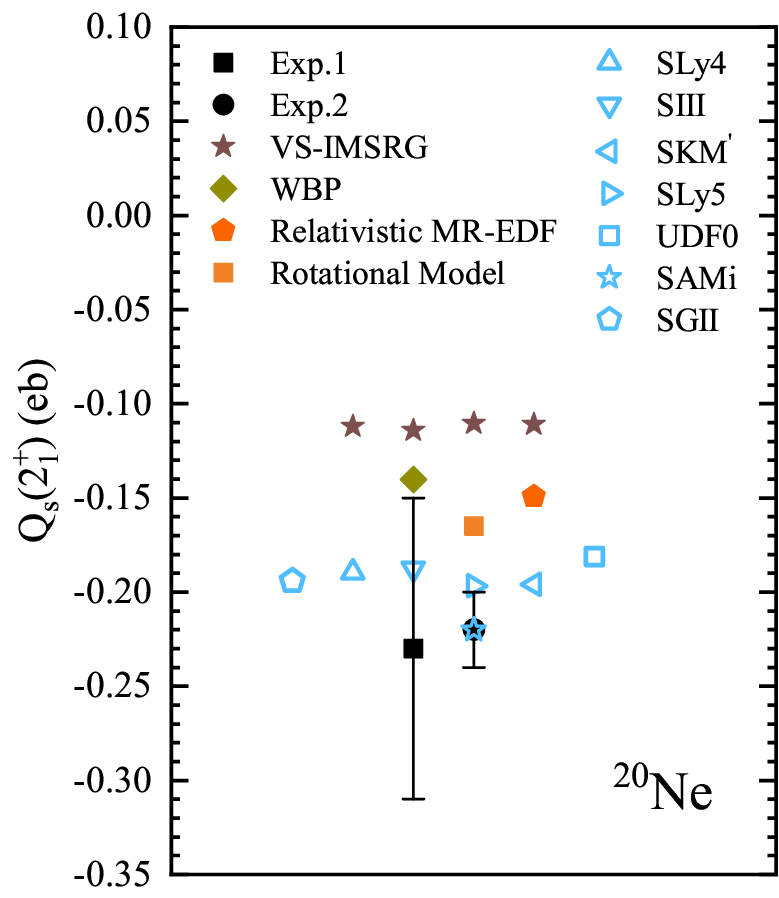}
  \caption{Experimental $Q_s(2_1^+)$ values for $^{20}$Ne 
  from Ref.~\cite{Schwalm1972NPA} (Exp.1) and 
  Ref.~\cite{mehl2024prc} (Exp.2), compared 
  with theoretical predictions from previous models 
  (VS-IMSRG, WBP, relativistic MR-EDF, and rotational 
  model)~\cite{mehl2024prc} and the current Beyond-SHF
  model. For VS-IMSRG results, interactions are 
  shown from left to right: PWA, N4LO, N2LO$_{\textrm{sat}}$, 
  and 1.8/2.0(EM).}\label{f:Q-compare}
\end{center}
\end{figure}

The obtained $Q_s(2_1^+)$ results by Beyond-SHF model are shown 
in Fig.~\ref{f:Q-compare} in comparisons with the experimental
data and the theoretical results presented in Ref.~\cite{mehl2024prc}. 
Our results indicate that Skyrme-type interactions generally 
yield $Q_s(2_1^+)$ values closer to experimental measurements 
than those of other models. Notably, the SAMi reproduces 
the experimental value exactly, yielding $Q_s(2_1^+)
= -0.22~e\textrm{b}$.

\begin{figure}[ht]
\begin{center}
  \includegraphics[width=\linewidth]{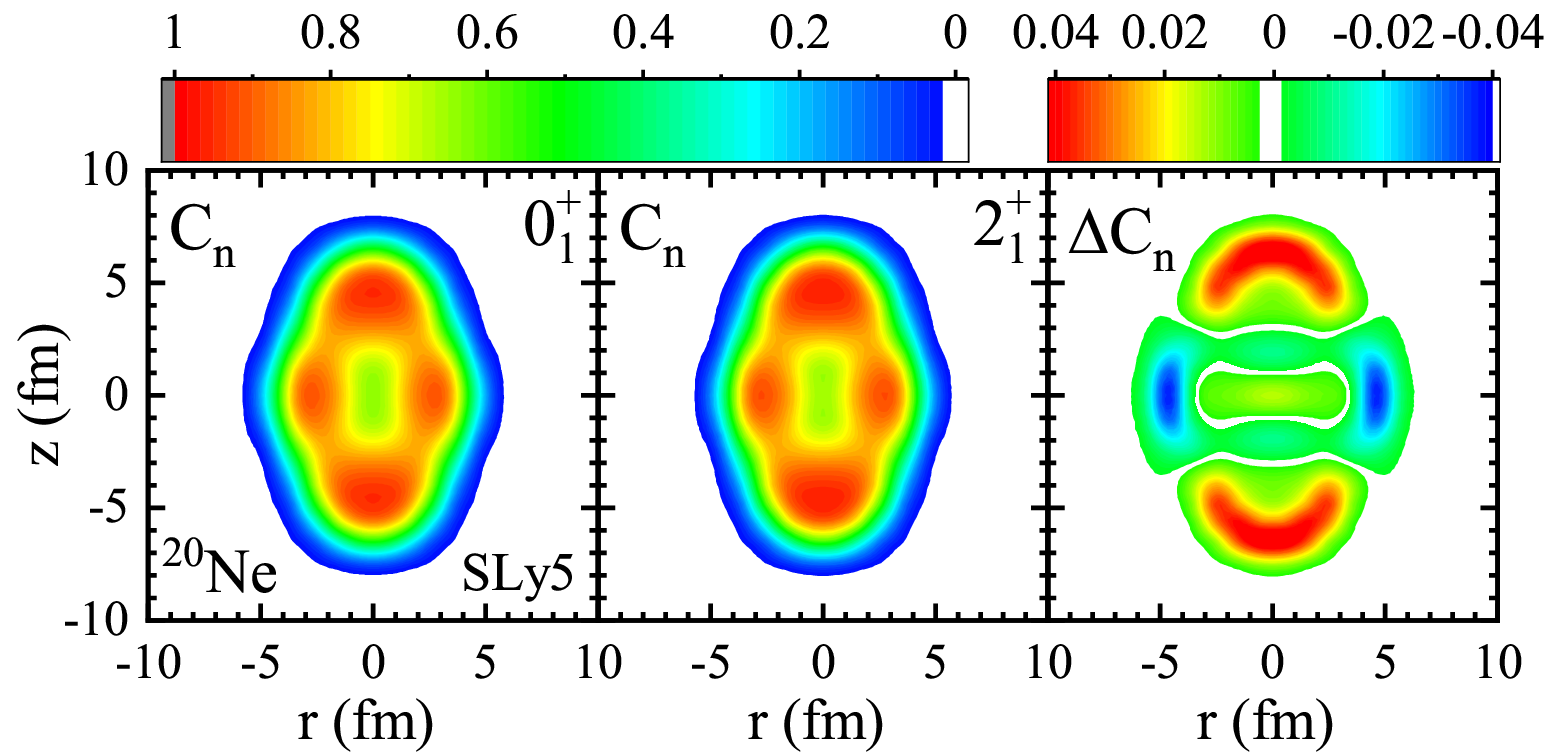}
  \caption{Left and middle: Localization functions
  $\mathcal{C}_n(0_1^+)$ for the ground state and 
  $\mathcal{C}_n(2_1^+)$ for the first excited state 
  of $^{20}$Ne, calculated using the SLy5 interaction. 
  Right: Difference $\Delta \mathcal{C}_n = \mathcal{C}_n(2_1^+) 
  - \mathcal{C}_n(0_1^+)$.}\label{f:loca}
\end{center}
\end{figure}

It is noted that the relativistic MR-EDF model in Ref.~\cite{mehl2024prc} 
reveals a ``bowling-pin" clustering structure in the $2_1^+$ state 
of $^{20}$Ne, characterized by octupole deformation. However, 
the current model employs an axisymmetric, parity-conserving model 
space, which cannot describe such octupole-deformed cluster 
structures. Nevertheless, it is still possible to investigate 
clustering or localized structures within a model space 
that accounts only the quadrupole deformation. To this end, 
we extend the localization function from the mean-field 
framework~\cite{C.L.Zhang2016prc, Schuetrumpf2017prc, Mercier2021prc, Li2024prc} to beyond-mean-field 
nuclear states, incorporating angular momentum projection 
and the generator coordinate method.

The localization functions for the ground state, $\mathcal{C}_n(0_1^+)$, 
and the first excited state, $\mathcal{C}_n(2_1^+)$, 
calculated with the SLy5 functional, are presented
in Fig.~\ref{f:loca}. Both functions reach a maximum value 
of $\approx 0.95$, indicating a pronounced $\alpha$ clustering 
structure in both states. Analysis of the difference between 
the localization functions, $\Delta \mathcal{C}_n
= \mathcal{C}_n(2_1^+) - \mathcal{C}_n(0_1^+)$, reveals 
that the $2_1^+$ state exhibits stronger clustering 
along the $z$ axis compared to the $0_1^+$ state. 
This suggests that the nuclear rotational motion 
along the $r$ direction enhances the clustering 
structure along the $z$ axis, likely due to the 
Coriolis interaction. These findings further 
corroborate the presence of $\alpha$ clustering 
in the $2_1^+$ state of $^{20}\textrm{Ne}$.

\begin{figure}[ht]
\begin{center}
  \includegraphics[width=0.95\linewidth]{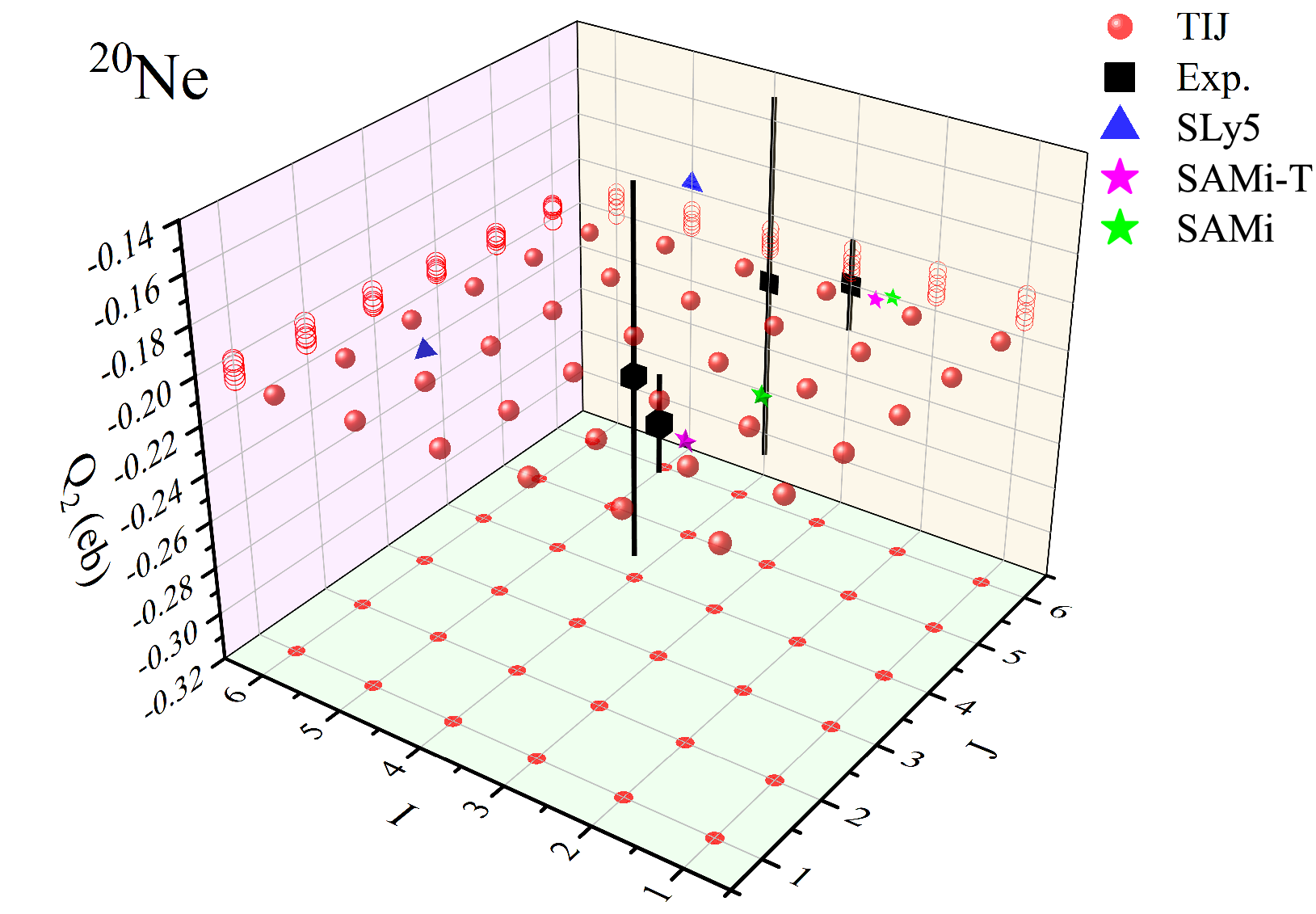}
  \caption{Comparison of experimental and theoretical 
  $Q_s(2_1^+)$ values for $^{20}$Ne, calculated by
  the Beyond-SHF model with TIJ  (ranging from T11 
  to T66) sets~\cite{TIJ}, SLy5, SAMi, and SAMi+T 
  interactions.}\label{f:Q-3d}
\end{center}
\end{figure}

In the following, the influence of tensor-forces on 
the $Q_s(2_1^+)$ of $^{20}$Ne is investigated. 
Fig.~\ref{f:Q-3d} compares the calculated $Q_s(2_1^+)$ 
values using the TIJ (ranging from T11 to T66) 
parameter sets~\cite{TIJ} with those obtained from 
SLy5, SAMi, and SAMi+T interactions. As anticipated, 
the inclusion of tensor-forces enhances the magnitude 
of $Q_s(2_1^+)$. A systematic increase in $Q_s(2_1^+)$ 
is observed with increasing tensor-force strength 
parameters $\alpha_T$ and $\beta_T$. 

As summarized in Table~\ref{t:qs}, calculations across the 
36 TIJ parameter sets yield $Q_s(2_1^+)$ values ranging 
from $-0.202~e\textrm{b}$ to $-0.221~e\textrm{b}$, 
demonstrating close agreement with experimental 
measurements. Furthermore, a comparison between 
SAMi ($-0.220~e\textrm{b}$) and SAMi+T ($-0.223~e\textrm{b}$) 
indicates that, while the absolute value of $Q_s(2_1^+)$ 
in SAMi+T is only marginally larger, tensor 
interactions consistently contribute to an 
enhanced $Q_s(2_1^+)$. These findings underscore 
the role of tensor-forces in mitigating 
the underestimation of $Q_s$ in nuclear
models, improving the alignment between 
theoretical predictions and experimental data.

\begin{table}[ht]
  \caption{Calculated $Q_s(2_1^+)$ values (in $e\textrm{b}$) 
  in $^{20}$Ne with $36$ sets of TIJ parameters.}\label{t:qs}
  \renewcommand\arraystretch{1.1}
  \setlength{\tabcolsep}{1.0 pt}
  \begin{ruledtabular}
  \begin{tabular}{ccc|ccc}
  I & J & $~~~~Q_s(2_1^+)~~~~$  & I & J & $~~~~Q_s(2_1^+)~~~~$\\
  \hline
  1 & 1 & $-0.202$      & 4 & 1 & $-0.206$\\
  1 & 2 & $-0.203$      & 4 & 2 & $-0.209$\\
  1 & 3 & $-0.206$      & 4 & 3 & $-0.211$\\
  1 & 4 & $-0.209$      & 4 & 4 & $-0.213$\\
  1 & 5 & $-0.211$      & 4 & 5 & $-0.214$\\
  1 & 6 & $-0.214$      & 4 & 6 & $-0.216$\\
  2 & 1 & $-0.203$      & 5 & 1 & $-0.208$\\
  2 & 2 & $-0.206$      & 5 & 2 & $-0.210$\\
  2 & 3 & $-0.208$      & 5 & 3 & $-0.212$\\
  2 & 4 & $-0.211$      & 5 & 4 & $-0.214$\\
  2 & 5 & $-0.213$      & 5 & 5 & $-0.215$\\
  2 & 6 & $-0.214$      & 5 & 6 & $-0.216$\\
  3 & 1 & $-0.204$      & 6 & 1 & $-0.210$\\
  3 & 2 & $-0.208$      & 6 & 2 & $-0.212$\\
  3 & 3 & $-0.210$      & 6 & 3 & $-0.213$\\
  3 & 4 & $-0.212$      & 6 & 4 & $-0.215$\\
  3 & 5 & $-0.214$      & 6 & 5 & $-0.217$\\
  3 & 6 & $-0.215$      & 6 & 6 & $-0.221$\\
  \end{tabular}
 \end{ruledtabular}
\end{table}

\begin{figure}[!h]
\begin{center}
  \includegraphics[width=0.80 \linewidth]{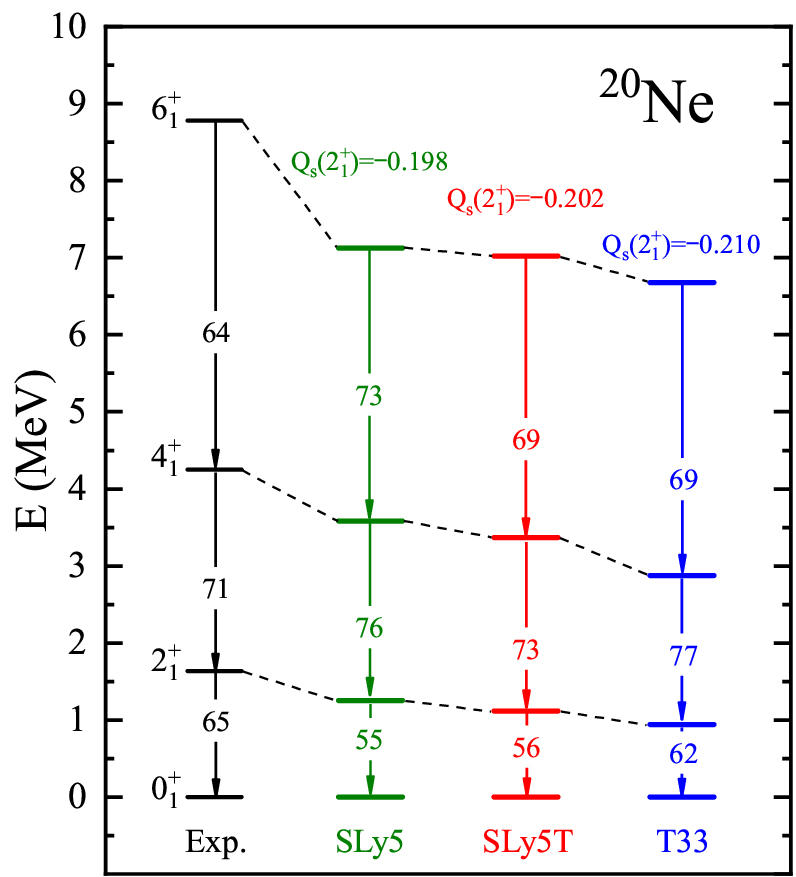}
  \caption{Experimental ground-state band of $^{20}$Ne 
  compared with theoretical predictions from the 
  Beyond-SHF model using SLy5, SLy5+T, 
  and T33 interactions. Theoretical $Q_s(2_1^+)$ 
  values (in $e\textrm{b}$) and intra-band $B(E2)$
  transition strengths (in $e^2\textrm{fm}^4$) 
  are also provided.}\label{f:fig4}
\end{center}
\end{figure}

To further elucidate the impact of tensor-force on 
nuclear structure, we conducted comparative analyses 
of the ground-state band structure in $^{20}$Ne using 
three parameter sets: SLy5 (without tensor-force), 
SLy5+T (with tensor-force), and T33 (with tensor-force). 
The T33 parametrization was chosen due to its excellent agreement 
with experimental quadrupole moment values and its intermediate 
tensor strength among the 36 Skyrme-type parameter sets, 
making it a representative benchmark for investigating 
generalized tensor interactions.

The calculated energy spectra of the ground-state band 
in $^{20}$Ne are presented in Fig.~\ref{f:fig4}, compared
with experimental data. The low-lying spectra obtained 
using the SLy5+T and T33 interactions closely resemble 
those from SLy5, but exhibit slightly reduced energy
spacings between levels. This suggests that tensor-force
enhances nuclear deformation, resulting in an increased 
moment of inertia. Consequently, the intra-band $B(E2)$ 
transition strengths are also enhanced.

\begin{figure}[ht]
\begin{center}
  \includegraphics[width=\linewidth]{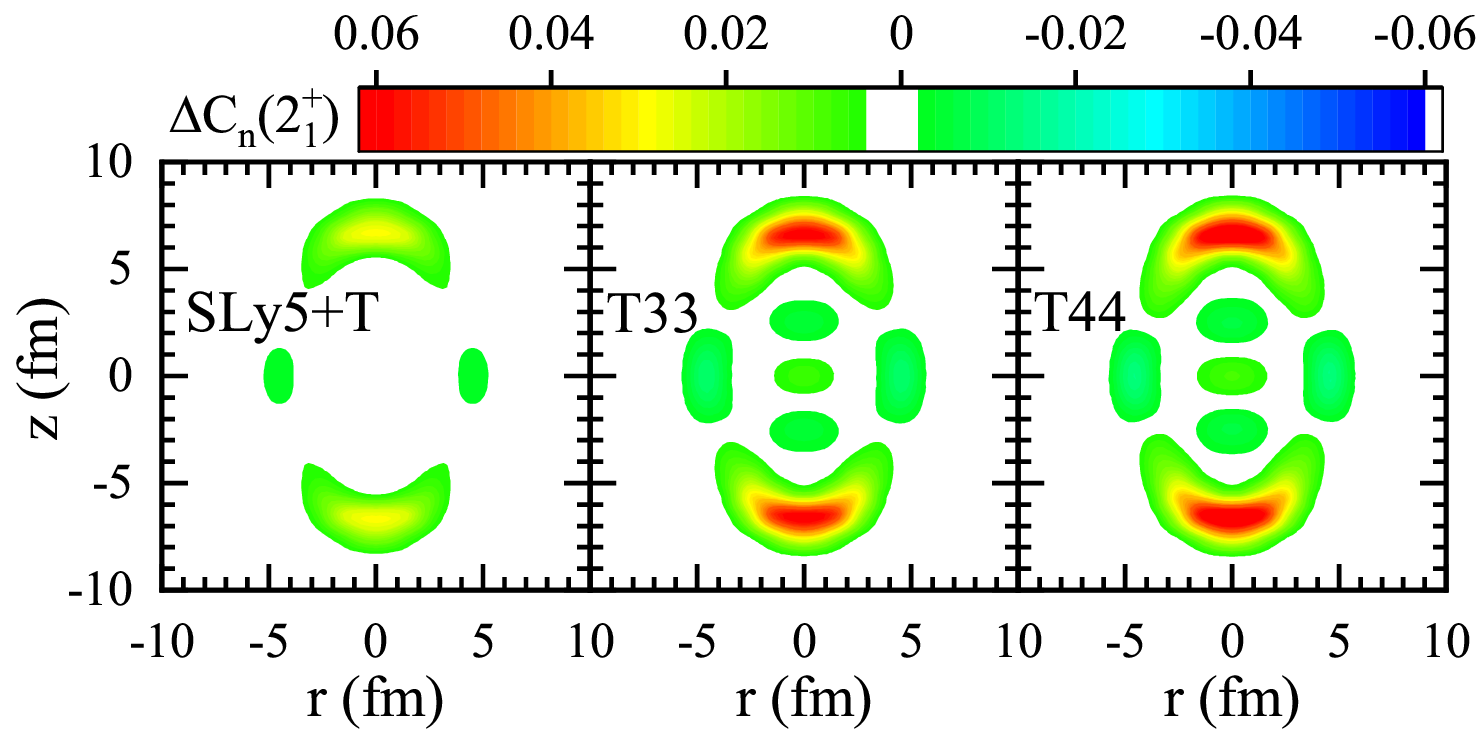}
  \caption{Localization function difference $\Delta \mathcal{C}_n$
  for the $2_1^+$ state of $^{20}$Ne, comparing 
  SLy5 with SLy5+T (left), T33 (middle), 
  and T44 (right).}\label{f:fig5}
\end{center}
\end{figure}

Therefore, we have confirmed that incorporating tensor 
forces into the interaction optimizes the theoretical 
model's predictive accuracy for the $Q_s(2_1^+)$. To 
further explore whether this improvement in $Q_s(2_1^+)$ 
is linked to enhanced cluster structures induced by 
tensor-force, we analyzed the localization function 
differences for the SLy5, T33, and T44 parameter 
sets, with and without tensor-force, as shown 
in Fig.~\ref{f:fig5}. The results clearly 
demonstrate that tensor interactions significantly 
alter the localization function, manifesting as 
an enhanced localization along the $z$ direction and 
a slight reduction along the $r$ direction. This 
indicates that tensor-force not only amplifies nuclear 
deformation but also increases the propensity for 
clustering in specific directions.

\section{Summary}
\label{s:summary}

In summary, we employ the Beyond-Skyrme-Hartree-Fock framework 
with various Skyrme-type nucleon-nucleon effective interactions 
to investigate the large spectroscopic quadrupole moment
$Q_s$ of the first excited $2_1^+$ state in $^{20}$Ne. 
Our calculated $Q_s(2_1^+) \approx -0.20~e\textrm{b}$ 
is in close agreement with the experimental value of 
$-0.22(2)~e\textrm{b}$. Consistent with relativistic
MR-EDF model, our findings confirm the presence of an 
$\alpha$ cluster structure in this state. Furthermore, 
we assess the stability of the $\alpha$ cluster structure 
under the influence of tensor-force, demonstrating 
that their inclusion significantly enhances $Q_s(2_1^+)$ 
and amplifies the clustering along the $z$ axis.
The influence of tensor-force on clustering structures 
in spatially reflection-asymmetric, octupole-deformed 
nuclei will be systematically investigated in future studies.

\section*{Acknowledgements}

This work was supported by the Doctoral Specialized Project 
of Nanyang Normal University~(No.~2025ZX006), the National 
Natural Science Foundation of China~(Grant No.~12175071,
No.~12205103, No.~12035011, and No.~11975167), 
the National Key R\&D Program of China~(No.~2024YFE0109803, No.~2023YFA1606503).

\newcommand{\epja}{EPJA}
\newcommand{\npa}{Nucl. Phys. A}
\newcommand{\nphysa}{Nucl. Phys. A}
\newcommand{\ppnp}{Prog. Part. Nucl. Phys.}
\newcommand{\ptp}{Prog. Theor. Phys.}
\newcommand{\ptep}{Prog. Theor. Exp. Phys.}

\bibliography{Ne-Qs}

\begin{thebibliography}{61}%
\makeatletter
\providecommand \@ifxundefined [1]{%
 \@ifx{#1\undefined}
}%
\providecommand \@ifnum [1]{%
 \ifnum #1\expandafter \@firstoftwo
 \else \expandafter \@secondoftwo
 \fi
}%
\providecommand \@ifx [1]{%
 \ifx #1\expandafter \@firstoftwo
 \else \expandafter \@secondoftwo
 \fi
}%
\providecommand \natexlab [1]{#1}%
\providecommand \enquote  [1]{``#1''}%
\providecommand \bibnamefont  [1]{#1}%
\providecommand \bibfnamefont [1]{#1}%
\providecommand \citenamefont [1]{#1}%
\providecommand \href@noop [0]{\@secondoftwo}%
\providecommand \href [0]{\begingroup \@sanitize@url \@href}%
\providecommand \@href[1]{\@@startlink{#1}\@@href}%
\providecommand \@@href[1]{\endgroup#1\@@endlink}%
\providecommand \@sanitize@url [0]{\catcode `\\12\catcode `\$12\catcode
  `\&12\catcode `\#12\catcode `\^12\catcode `\_12\catcode `\%12\relax}%
\providecommand \@@startlink[1]{}%
\providecommand \@@endlink[0]{}%
\providecommand \url  [0]{\begingroup\@sanitize@url \@url }%
\providecommand \@url [1]{\endgroup\@href {#1}{\urlprefix }}%
\providecommand \urlprefix  [0]{URL }%
\providecommand \Eprint [0]{\href }%
\providecommand \doibase [0]{http://dx.doi.org/}%
\providecommand \selectlanguage [0]{\@gobble}%
\providecommand \bibinfo  [0]{\@secondoftwo}%
\providecommand \bibfield  [0]{\@secondoftwo}%
\providecommand \translation [1]{[#1]}%
\providecommand \BibitemOpen [0]{}%
\providecommand \bibitemStop [0]{}%
\providecommand \bibitemNoStop [0]{.\EOS\space}%
\providecommand \EOS [0]{\spacefactor3000\relax}%
\providecommand \BibitemShut  [1]{\csname bibitem#1\endcsname}%
\let\auto@bib@innerbib\@empty
\bibitem [{\citenamefont {Carchidi}\ \emph {et~al.}(1986)\citenamefont
  {Carchidi}, \citenamefont {Wildenthal},\ and\ \citenamefont
  {Brown}}]{Carchidi1986prc}%
  \BibitemOpen
  \bibfield  {author} {\bibinfo {author} {\bibfnamefont {M.}~\bibnamefont
  {Carchidi}}, \bibinfo {author} {\bibfnamefont {B.~H.}\ \bibnamefont
  {Wildenthal}}, \ and\ \bibinfo {author} {\bibfnamefont {B.~A.}\ \bibnamefont
  {Brown}},\ }\href {\doibase 10.1103/PhysRevC.34.2280} {\bibfield  {journal}
  {\bibinfo  {journal} {Phys. Rev. C}\ }\textbf {\bibinfo {volume} {34}},\
  \bibinfo {pages} {2280} (\bibinfo {year} {1986})}\BibitemShut {NoStop}%
\bibitem [{\citenamefont {Neyens}(2003)}]{Neyens2003RPP}%
  \BibitemOpen
  \bibfield  {author} {\bibinfo {author} {\bibfnamefont {G.}~\bibnamefont
  {Neyens}},\ }\href@noop {} {\bibfield  {journal} {\bibinfo  {journal} {Rep.
  Prog. Phys.}\ }\textbf {\bibinfo {volume} {66}},\ \bibinfo {pages} {633}
  (\bibinfo {year} {2003})}\BibitemShut {NoStop}%
\bibitem [{\citenamefont {Lederer}\ and\ \citenamefont
  {Shirley}(1978)}]{Lederer1978tableof}%
  \BibitemOpen
  \bibfield  {author} {\bibinfo {author} {\bibfnamefont {C.~M.}\ \bibnamefont
  {Lederer}}\ and\ \bibinfo {author} {\bibfnamefont {V.~S.}\ \bibnamefont
  {Shirley}},\ }\href@noop {} {\enquote {\bibinfo {title} {Tableof isotopes,
  seventhedition},}\ } (\bibinfo {year} {1978})\BibitemShut {NoStop}%
\bibitem [{\citenamefont {Raghavan}(1989)}]{Raghavan1989ADNDT}%
  \BibitemOpen
  \bibfield  {author} {\bibinfo {author} {\bibfnamefont {P.}~\bibnamefont
  {Raghavan}},\ }\href {\doibase https://doi.org/10.1016/0092-640X(89)90008-9}
  {\bibfield  {journal} {\bibinfo  {journal} {Atom. Data Nucl. Data}\ }\textbf
  {\bibinfo {volume} {42}},\ \bibinfo {pages} {189} (\bibinfo {year}
  {1989})}\BibitemShut {NoStop}%
\bibitem [{\citenamefont {Stone}(2005)}]{Stone2005ADNDT}%
  \BibitemOpen
  \bibfield  {author} {\bibinfo {author} {\bibfnamefont {N.}~\bibnamefont
  {Stone}},\ }\href {\doibase https://doi.org/10.1016/j.adt.2005.04.001}
  {\bibfield  {journal} {\bibinfo  {journal} {Atom. Data Nucl. Data}\ }\textbf
  {\bibinfo {volume} {90}},\ \bibinfo {pages} {75} (\bibinfo {year}
  {2005})}\BibitemShut {NoStop}%
\bibitem [{\citenamefont {Loebner}\ \emph {et~al.}(1970)\citenamefont
  {Loebner}, \citenamefont {Vetter},\ and\ \citenamefont
  {Hoenig}}]{Loebner1970NDSA}%
  \BibitemOpen
  \bibfield  {author} {\bibinfo {author} {\bibfnamefont {K.~E.}\ \bibnamefont
  {Loebner}}, \bibinfo {author} {\bibfnamefont {M.}~\bibnamefont {Vetter}}, \
  and\ \bibinfo {author} {\bibfnamefont {V.}~\bibnamefont {Hoenig}},\ }\href
  {https://www.osti.gov/biblio/4109385} {\bibfield  {journal} {\bibinfo
  {journal} {Nucl. Data, Sect. A}\ }\textbf {\bibinfo {volume} {7}},\ \bibinfo
  {pages} {495} (\bibinfo {year} {1970})}\BibitemShut {NoStop}%
\bibitem [{\citenamefont {Christy}\ and\ \citenamefont
  {Haeusser}(1973)}]{Christy1973NDSA}%
  \BibitemOpen
  \bibfield  {author} {\bibinfo {author} {\bibfnamefont {A.}~\bibnamefont
  {Christy}}\ and\ \bibinfo {author} {\bibfnamefont {O.}~\bibnamefont
  {Haeusser}},\ }\href {\doibase 10.1016/S0092-640X(73)80018-X} {\bibfield
  {journal} {\bibinfo  {journal} {Nucl. Data, Sect. A}\ }\textbf {\bibinfo
  {volume} {11}},\ \bibinfo {pages} {281} (\bibinfo {year} {1973})}\BibitemShut
  {NoStop}%
\bibitem [{\citenamefont {Spear}(1981)}]{Spear1981PR}%
  \BibitemOpen
  \bibfield  {author} {\bibinfo {author} {\bibfnamefont {R.}~\bibnamefont
  {Spear}},\ }\href {\doibase https://doi.org/10.1016/0370-1573(81)90177-0}
  {\bibfield  {journal} {\bibinfo  {journal} {Phys. Rep.}\ }\textbf {\bibinfo
  {volume} {73}},\ \bibinfo {pages} {369} (\bibinfo {year} {1981})}\BibitemShut
  {NoStop}%
\bibitem [{\citenamefont {Mehl}\ \emph {et~al.}(2025)\citenamefont {Mehl},
  \citenamefont {Orce}, \citenamefont {Ngwetsheni}, \citenamefont
  {Marevi\ifmmode~\acute{c}\else \'{c}\fi{}}, \citenamefont {Brown},
  \citenamefont {Holt}, \citenamefont {Kumar~Raju}, \citenamefont {Lawrie},
  \citenamefont {Abrahams}, \citenamefont {Adsley}, \citenamefont {Akakpo},
  \citenamefont {Bark}, \citenamefont {Bernier}, \citenamefont {Bucher},
  \citenamefont {Yahia-Cherif}, \citenamefont {Dinoko}, \citenamefont {Ebran},
  \citenamefont {Erasmus}, \citenamefont {Jones}, \citenamefont {Khan},
  \citenamefont {Kheswa}, \citenamefont {Khumalo}, \citenamefont {Lawrie},
  \citenamefont {Majola}, \citenamefont {Malatji}, \citenamefont {Mavela},
  \citenamefont {Mokgolobotho}, \citenamefont {Nik\ifmmode \check{s}\else
  \v{s}\fi{}i\ifmmode~\acute{c}\else \'{c}\fi{}}, \citenamefont {Ntshangase},
  \citenamefont {Pesudo}, \citenamefont {Rebeiro}, \citenamefont {Shirinda},
  \citenamefont {Vretenar},\ and\ \citenamefont {Wiedeking}}]{mehl2024prc}%
  \BibitemOpen
  \bibfield  {author} {\bibinfo {author} {\bibfnamefont {C.~V.}\ \bibnamefont
  {Mehl}}, \bibinfo {author} {\bibfnamefont {J.~N.}\ \bibnamefont {Orce}},
  \bibinfo {author} {\bibfnamefont {C.}~\bibnamefont {Ngwetsheni}}, \bibinfo
  {author} {\bibfnamefont {P.}~\bibnamefont {Marevi\ifmmode~\acute{c}\else
  \'{c}\fi{}}}, \bibinfo {author} {\bibfnamefont {B.~A.}\ \bibnamefont
  {Brown}}, \bibinfo {author} {\bibfnamefont {J.~D.}\ \bibnamefont {Holt}},
  \bibinfo {author} {\bibfnamefont {M.}~\bibnamefont {Kumar~Raju}}, \bibinfo
  {author} {\bibfnamefont {E.~A.}\ \bibnamefont {Lawrie}}, \bibinfo {author}
  {\bibfnamefont {K.~J.}\ \bibnamefont {Abrahams}}, \bibinfo {author}
  {\bibfnamefont {P.}~\bibnamefont {Adsley}}, \bibinfo {author} {\bibfnamefont
  {E.~H.}\ \bibnamefont {Akakpo}}, \bibinfo {author} {\bibfnamefont {R.~A.}\
  \bibnamefont {Bark}}, \bibinfo {author} {\bibfnamefont {N.}~\bibnamefont
  {Bernier}}, \bibinfo {author} {\bibfnamefont {T.~D.}\ \bibnamefont {Bucher}},
  \bibinfo {author} {\bibfnamefont {W.}~\bibnamefont {Yahia-Cherif}}, \bibinfo
  {author} {\bibfnamefont {T.~S.}\ \bibnamefont {Dinoko}}, \bibinfo {author}
  {\bibfnamefont {J.-P.}\ \bibnamefont {Ebran}}, \bibinfo {author}
  {\bibfnamefont {N.}~\bibnamefont {Erasmus}}, \bibinfo {author} {\bibfnamefont
  {P.~M.}\ \bibnamefont {Jones}}, \bibinfo {author} {\bibfnamefont
  {E.}~\bibnamefont {Khan}}, \bibinfo {author} {\bibfnamefont {N.~Y.}\
  \bibnamefont {Kheswa}}, \bibinfo {author} {\bibfnamefont {N.~A.}\
  \bibnamefont {Khumalo}}, \bibinfo {author} {\bibfnamefont {J.~J.}\
  \bibnamefont {Lawrie}}, \bibinfo {author} {\bibfnamefont {S.~N.~T.}\
  \bibnamefont {Majola}}, \bibinfo {author} {\bibfnamefont {K.~L.}\
  \bibnamefont {Malatji}}, \bibinfo {author} {\bibfnamefont {D.~L.}\
  \bibnamefont {Mavela}}, \bibinfo {author} {\bibfnamefont {M.~J.}\
  \bibnamefont {Mokgolobotho}}, \bibinfo {author} {\bibfnamefont
  {T.}~\bibnamefont {Nik\ifmmode \check{s}\else
  \v{s}\fi{}i\ifmmode~\acute{c}\else \'{c}\fi{}}}, \bibinfo {author}
  {\bibfnamefont {S.~S.}\ \bibnamefont {Ntshangase}}, \bibinfo {author}
  {\bibfnamefont {V.}~\bibnamefont {Pesudo}}, \bibinfo {author} {\bibfnamefont
  {B.}~\bibnamefont {Rebeiro}}, \bibinfo {author} {\bibfnamefont
  {O.}~\bibnamefont {Shirinda}}, \bibinfo {author} {\bibfnamefont
  {D.}~\bibnamefont {Vretenar}}, \ and\ \bibinfo {author} {\bibfnamefont
  {M.}~\bibnamefont {Wiedeking}},\ }\href {\doibase
  10.1103/PhysRevC.111.054318} {\bibfield  {journal} {\bibinfo  {journal}
  {Phys. Rev. C}\ }\textbf {\bibinfo {volume} {111}},\ \bibinfo {pages}
  {054318} (\bibinfo {year} {2025})}\BibitemShut {NoStop}%
\bibitem [{\citenamefont {Aaij}\ \emph {et~al.}(2022)\citenamefont {Aaij},
  \citenamefont {Beteta}, \citenamefont {Ackernley}, \citenamefont {Adeva},
  \citenamefont {Adinolfi}, \citenamefont {Afsharnia}, \citenamefont {Aidala},
  \citenamefont {Aiola}, \citenamefont {Ajaltouni}, \citenamefont {Akar} \emph
  {et~al.}}]{Aaij2022JoI}%
  \BibitemOpen
  \bibfield  {author} {\bibinfo {author} {\bibfnamefont {R.}~\bibnamefont
  {Aaij}}, \bibinfo {author} {\bibfnamefont {C.~A.}\ \bibnamefont {Beteta}},
  \bibinfo {author} {\bibfnamefont {T.}~\bibnamefont {Ackernley}}, \bibinfo
  {author} {\bibfnamefont {B.}~\bibnamefont {Adeva}}, \bibinfo {author}
  {\bibfnamefont {M.}~\bibnamefont {Adinolfi}}, \bibinfo {author}
  {\bibfnamefont {H.}~\bibnamefont {Afsharnia}}, \bibinfo {author}
  {\bibfnamefont {C.~A.}\ \bibnamefont {Aidala}}, \bibinfo {author}
  {\bibfnamefont {S.}~\bibnamefont {Aiola}}, \bibinfo {author} {\bibfnamefont
  {Z.}~\bibnamefont {Ajaltouni}}, \bibinfo {author} {\bibfnamefont
  {S.}~\bibnamefont {Akar}},  \emph {et~al.},\ }\href@noop {} {\bibfield
  {journal} {\bibinfo  {journal} {J. Instrum.}\ }\textbf {\bibinfo {volume}
  {17}},\ \bibinfo {pages} {P05009} (\bibinfo {year} {2022})}\BibitemShut
  {NoStop}%
\bibitem [{\citenamefont {Ebran}\ \emph {et~al.}(2012)\citenamefont {Ebran},
  \citenamefont {Khan}, \citenamefont {Nik{\v{s}}i{\'c}},\ and\ \citenamefont
  {Vretenar}}]{Ebran2012nature}%
  \BibitemOpen
  \bibfield  {author} {\bibinfo {author} {\bibfnamefont {J.-P.}\ \bibnamefont
  {Ebran}}, \bibinfo {author} {\bibfnamefont {E.}~\bibnamefont {Khan}},
  \bibinfo {author} {\bibfnamefont {T.}~\bibnamefont {Nik{\v{s}}i{\'c}}}, \
  and\ \bibinfo {author} {\bibfnamefont {D.}~\bibnamefont {Vretenar}},\ }\href
  {\doibase 10.1038/nature11246} {\bibfield  {journal} {\bibinfo  {journal}
  {Nature}\ }\textbf {\bibinfo {volume} {487}},\ \bibinfo {pages} {341}
  (\bibinfo {year} {2012})}\BibitemShut {NoStop}%
\bibitem [{\citenamefont {M\"antysaari}\ \emph {et~al.}(2023)\citenamefont
  {M\"antysaari}, \citenamefont {Schenke}, \citenamefont {Shen},\ and\
  \citenamefont {Zhao}}]{Mantysaari2023PRL}%
  \BibitemOpen
  \bibfield  {author} {\bibinfo {author} {\bibfnamefont {H.}~\bibnamefont
  {M\"antysaari}}, \bibinfo {author} {\bibfnamefont {B.}~\bibnamefont
  {Schenke}}, \bibinfo {author} {\bibfnamefont {C.}~\bibnamefont {Shen}}, \
  and\ \bibinfo {author} {\bibfnamefont {W.}~\bibnamefont {Zhao}},\ }\href
  {\doibase 10.1103/PhysRevLett.131.062301} {\bibfield  {journal} {\bibinfo
  {journal} {Phys. Rev. Lett.}\ }\textbf {\bibinfo {volume} {131}},\ \bibinfo
  {pages} {062301} (\bibinfo {year} {2023})}\BibitemShut {NoStop}%
\bibitem [{\citenamefont {Nakai}\ \emph {et~al.}(1970)\citenamefont {Nakai},
  \citenamefont {Stephens},\ and\ \citenamefont {Diamond}}]{Nakai1970NPA}%
  \BibitemOpen
  \bibfield  {author} {\bibinfo {author} {\bibfnamefont {K.}~\bibnamefont
  {Nakai}}, \bibinfo {author} {\bibfnamefont {F.~S.}\ \bibnamefont {Stephens}},
  \ and\ \bibinfo {author} {\bibfnamefont {R.~M.}\ \bibnamefont {Diamond}},\
  }\href@noop {} {\bibfield  {journal} {\bibinfo  {journal} {Nucl. Phys. A}\
  }\textbf {\bibinfo {volume} {150}},\ \bibinfo {pages} {114} (\bibinfo {year}
  {1970})}\BibitemShut {NoStop}%
\bibitem [{\citenamefont {Schwalm}\ \emph {et~al.}(1972)\citenamefont
  {Schwalm}, \citenamefont {Bamberger}, \citenamefont {Bizzeti}, \citenamefont
  {Povh}, \citenamefont {Engelbertink}, \citenamefont {Olness},\ and\
  \citenamefont {Warburton}}]{Schwalm1972NPA}%
  \BibitemOpen
  \bibfield  {author} {\bibinfo {author} {\bibfnamefont {D.}~\bibnamefont
  {Schwalm}}, \bibinfo {author} {\bibfnamefont {A.}~\bibnamefont {Bamberger}},
  \bibinfo {author} {\bibfnamefont {P.~G.}\ \bibnamefont {Bizzeti}}, \bibinfo
  {author} {\bibfnamefont {B.}~\bibnamefont {Povh}}, \bibinfo {author}
  {\bibfnamefont {G.~A.~P.}\ \bibnamefont {Engelbertink}}, \bibinfo {author}
  {\bibfnamefont {J.~W.}\ \bibnamefont {Olness}}, \ and\ \bibinfo {author}
  {\bibfnamefont {E.~K.}\ \bibnamefont {Warburton}},\ }\href@noop {} {\bibfield
   {journal} {\bibinfo  {journal} {Nucl. Phys. A}\ }\textbf {\bibinfo {volume}
  {192}},\ \bibinfo {pages} {449} (\bibinfo {year} {1972})}\BibitemShut
  {NoStop}%
\bibitem [{\citenamefont {Olsen}\ \emph {et~al.}(1974)\citenamefont {Olsen},
  \citenamefont {Barnett}, \citenamefont {Biagi}, \citenamefont {Merrill},\
  and\ \citenamefont {Phillips}}]{Olsen1974NPA}%
  \BibitemOpen
  \bibfield  {author} {\bibinfo {author} {\bibfnamefont {D.~K.}\ \bibnamefont
  {Olsen}}, \bibinfo {author} {\bibfnamefont {A.~R.}\ \bibnamefont {Barnett}},
  \bibinfo {author} {\bibfnamefont {S.~F.}\ \bibnamefont {Biagi}}, \bibinfo
  {author} {\bibfnamefont {N.~H.}\ \bibnamefont {Merrill}}, \ and\ \bibinfo
  {author} {\bibfnamefont {W.~R.}\ \bibnamefont {Phillips}},\ }\href@noop {}
  {\bibfield  {journal} {\bibinfo  {journal} {Nucl. Phys. A}\ }\textbf
  {\bibinfo {volume} {220}},\ \bibinfo {pages} {541} (\bibinfo {year}
  {1974})}\BibitemShut {NoStop}%
\bibitem [{NuD()}]{NuDat}%
  \BibitemOpen
  \href@noop {} {\enquote {\bibinfo {title} {National nuclear data center},}\
  }\bibinfo {note} {\url{https://www.nndc.bnl.gov/nudat3/}}\BibitemShut
  {NoStop}%
\bibitem [{\citenamefont {Bohr}\ and\ \citenamefont {Mottelson}(1998)}]{bm}%
  \BibitemOpen
  \bibfield  {author} {\bibinfo {author} {\bibfnamefont {A.}~\bibnamefont
  {Bohr}}\ and\ \bibinfo {author} {\bibfnamefont {B.~R.}\ \bibnamefont
  {Mottelson}},\ }\href@noop {} {\emph {\bibinfo {title} {Nuclear Structure V.
  II}}}\ (\bibinfo  {publisher} {World Scientific Publishing Company},\
  \bibinfo {year} {1998})\BibitemShut {NoStop}%
\bibitem [{\citenamefont {Thomson}(1913)}]{thomson1913bakerian}%
  \BibitemOpen
  \bibfield  {author} {\bibinfo {author} {\bibfnamefont {J.~J.}\ \bibnamefont
  {Thomson}},\ }\href@noop {} {\bibfield  {journal} {\bibinfo  {journal} {Proc.
  Royal. Society A}\ }\textbf {\bibinfo {volume} {89}},\ \bibinfo {pages} {1}
  (\bibinfo {year} {1913})}\BibitemShut {NoStop}%
\bibitem [{\citenamefont {Rodr\'{i}guez-Guzm\'{a}n}\ \emph
  {et~al.}(2003)\citenamefont {Rodr\'{i}guez-Guzm\'{a}n}, \citenamefont
  {Egido},\ and\ \citenamefont {Robledo}}]{20Negogny}%
  \BibitemOpen
  \bibfield  {author} {\bibinfo {author} {\bibfnamefont {R.~R.}\ \bibnamefont
  {Rodr\'{i}guez-Guzm\'{a}n}}, \bibinfo {author} {\bibfnamefont {J.~L.}\
  \bibnamefont {Egido}}, \ and\ \bibinfo {author} {\bibfnamefont {L.~M.}\
  \bibnamefont {Robledo}},\ }\href {\doibase 10.1140/epja/i2002-10141-6}
  {\bibfield  {journal} {\bibinfo  {journal} {Eur. Phys. J. A.}\ }\textbf
  {\bibinfo {volume} {17}},\ \bibinfo {pages} {37–47} (\bibinfo {year}
  {2003})}\BibitemShut {NoStop}%
\bibitem [{\citenamefont {Zhou}\ \emph {et~al.}(2016)\citenamefont {Zhou},
  \citenamefont {Yao}, \citenamefont {Li}, \citenamefont {Meng},\ and\
  \citenamefont {Ring}}]{EFzhou2016PLB}%
  \BibitemOpen
  \bibfield  {author} {\bibinfo {author} {\bibfnamefont {E.~F.}\ \bibnamefont
  {Zhou}}, \bibinfo {author} {\bibfnamefont {J.~M.}\ \bibnamefont {Yao}},
  \bibinfo {author} {\bibfnamefont {Z.~P.}\ \bibnamefont {Li}}, \bibinfo
  {author} {\bibfnamefont {J.}~\bibnamefont {Meng}}, \ and\ \bibinfo {author}
  {\bibfnamefont {P.}~\bibnamefont {Ring}},\ }\href {\doibase
  https://doi.org/10.1016/j.physletb.2015.12.028} {\bibfield  {journal}
  {\bibinfo  {journal} {Phys. Lett. B}\ }\textbf {\bibinfo {volume} {753}},\
  \bibinfo {pages} {227} (\bibinfo {year} {2016})}\BibitemShut {NoStop}%
\bibitem [{\citenamefont {Marevi{\'c}}\ \emph {et~al.}(2018)\citenamefont
  {Marevi{\'c}}, \citenamefont {Ebran}, \citenamefont {Khan}, \citenamefont
  {Nik{\v{s}}i{\'c}},\ and\ \citenamefont {Vretenar}}]{marevic2018prc}%
  \BibitemOpen
  \bibfield  {author} {\bibinfo {author} {\bibfnamefont {P.}~\bibnamefont
  {Marevi{\'c}}}, \bibinfo {author} {\bibfnamefont {J.~P.}\ \bibnamefont
  {Ebran}}, \bibinfo {author} {\bibfnamefont {E.}~\bibnamefont {Khan}},
  \bibinfo {author} {\bibfnamefont {T.}~\bibnamefont {Nik{\v{s}}i{\'c}}}, \
  and\ \bibinfo {author} {\bibfnamefont {D.}~\bibnamefont {Vretenar}},\
  }\href@noop {} {\bibfield  {journal} {\bibinfo  {journal} {Phys. Rev. C}\
  }\textbf {\bibinfo {volume} {97}},\ \bibinfo {pages} {024334} (\bibinfo
  {year} {2018})}\BibitemShut {NoStop}%
\bibitem [{\citenamefont {Matsuse}\ \emph {et~al.}(1975)\citenamefont
  {Matsuse}, \citenamefont {Kamimura},\ and\ \citenamefont
  {Fukushima}}]{matsuse1975PTP}%
  \BibitemOpen
  \bibfield  {author} {\bibinfo {author} {\bibfnamefont {T.}~\bibnamefont
  {Matsuse}}, \bibinfo {author} {\bibfnamefont {M.}~\bibnamefont {Kamimura}}, \
  and\ \bibinfo {author} {\bibfnamefont {Y.}~\bibnamefont {Fukushima}},\ }\href
  {\doibase 10.1143/PTP.53.706} {\bibfield  {journal} {\bibinfo  {journal}
  {Prog. Theor. Phys.}\ }\textbf {\bibinfo {volume} {53}},\ \bibinfo {pages}
  {706} (\bibinfo {year} {1975})}\BibitemShut {NoStop}%
\bibitem [{\citenamefont {Le~Bloas}\ \emph {et~al.}(2014)\citenamefont
  {Le~Bloas}, \citenamefont {Pillet}, \citenamefont {Dupuis}, \citenamefont
  {Daugas}, \citenamefont {Robledo}, \citenamefont {Robin},\ and\ \citenamefont
  {Zelevinsky}}]{lebloas2014prc}%
  \BibitemOpen
  \bibfield  {author} {\bibinfo {author} {\bibfnamefont {J.}~\bibnamefont
  {Le~Bloas}}, \bibinfo {author} {\bibfnamefont {N.}~\bibnamefont {Pillet}},
  \bibinfo {author} {\bibfnamefont {M.}~\bibnamefont {Dupuis}}, \bibinfo
  {author} {\bibfnamefont {J.~M.}\ \bibnamefont {Daugas}}, \bibinfo {author}
  {\bibfnamefont {L.~M.}\ \bibnamefont {Robledo}}, \bibinfo {author}
  {\bibfnamefont {C.}~\bibnamefont {Robin}}, \ and\ \bibinfo {author}
  {\bibfnamefont {V.~G.}\ \bibnamefont {Zelevinsky}},\ }\href {\doibase
  10.1103/PhysRevC.89.011306} {\bibfield  {journal} {\bibinfo  {journal} {Phys.
  Rev. C}\ }\textbf {\bibinfo {volume} {89}},\ \bibinfo {pages} {011306}
  (\bibinfo {year} {2014})}\BibitemShut {NoStop}%
\bibitem [{\citenamefont {Hergert}\ \emph {et~al.}(2016)\citenamefont
  {Hergert}, \citenamefont {Bogner}, \citenamefont {Morris}, \citenamefont
  {Schwenk},\ and\ \citenamefont {Tsukiyama}}]{Hergert2016PR}%
  \BibitemOpen
  \bibfield  {author} {\bibinfo {author} {\bibfnamefont {H.}~\bibnamefont
  {Hergert}}, \bibinfo {author} {\bibfnamefont {S.}~\bibnamefont {Bogner}},
  \bibinfo {author} {\bibfnamefont {T.}~\bibnamefont {Morris}}, \bibinfo
  {author} {\bibfnamefont {A.}~\bibnamefont {Schwenk}}, \ and\ \bibinfo
  {author} {\bibfnamefont {K.}~\bibnamefont {Tsukiyama}},\ }\href {\doibase
  https://doi.org/10.1016/j.physrep.2015.12.007} {\bibfield  {journal}
  {\bibinfo  {journal} {Phys. Rep.}\ }\textbf {\bibinfo {volume} {621}},\
  \bibinfo {pages} {165} (\bibinfo {year} {2016})}\BibitemShut {NoStop}%
\bibitem [{\citenamefont {Stroberg}\ \emph {et~al.}(2019)\citenamefont
  {Stroberg}, \citenamefont {Hergert}, \citenamefont {Bogner},\ and\
  \citenamefont {Holt}}]{Stroberg2019AR}%
  \BibitemOpen
  \bibfield  {author} {\bibinfo {author} {\bibfnamefont {S.~R.}\ \bibnamefont
  {Stroberg}}, \bibinfo {author} {\bibfnamefont {H.}~\bibnamefont {Hergert}},
  \bibinfo {author} {\bibfnamefont {S.~K.}\ \bibnamefont {Bogner}}, \ and\
  \bibinfo {author} {\bibfnamefont {J.~D.}\ \bibnamefont {Holt}},\ }\href@noop
  {} {\bibfield  {journal} {\bibinfo  {journal} {Annu. Rev. Nucl. Part. S}\
  }\textbf {\bibinfo {volume} {69}},\ \bibinfo {pages} {307} (\bibinfo {year}
  {2019})}\BibitemShut {NoStop}%
\bibitem [{\citenamefont {Bogner}\ \emph {et~al.}(2014)\citenamefont {Bogner},
  \citenamefont {Hergert}, \citenamefont {Holt}, \citenamefont {Schwenk},
  \citenamefont {Binder}, \citenamefont {Calci}, \citenamefont {Langhammer},\
  and\ \citenamefont {Roth}}]{Bogner2014PRL}%
  \BibitemOpen
  \bibfield  {author} {\bibinfo {author} {\bibfnamefont {S.~K.}\ \bibnamefont
  {Bogner}}, \bibinfo {author} {\bibfnamefont {H.}~\bibnamefont {Hergert}},
  \bibinfo {author} {\bibfnamefont {J.~D.}\ \bibnamefont {Holt}}, \bibinfo
  {author} {\bibfnamefont {A.}~\bibnamefont {Schwenk}}, \bibinfo {author}
  {\bibfnamefont {S.}~\bibnamefont {Binder}}, \bibinfo {author} {\bibfnamefont
  {A.}~\bibnamefont {Calci}}, \bibinfo {author} {\bibfnamefont
  {J.}~\bibnamefont {Langhammer}}, \ and\ \bibinfo {author} {\bibfnamefont
  {R.}~\bibnamefont {Roth}},\ }\href {\doibase 10.1103/PhysRevLett.113.142501}
  {\bibfield  {journal} {\bibinfo  {journal} {Phys. Rev. Lett.}\ }\textbf
  {\bibinfo {volume} {113}},\ \bibinfo {pages} {142501} (\bibinfo {year}
  {2014})}\BibitemShut {NoStop}%
\bibitem [{\citenamefont {Bonche}\ \emph {et~al.}(1990)\citenamefont {Bonche},
  \citenamefont {Dobaczewski}, \citenamefont {Flocard}, \citenamefont
  {Heenen},\ and\ \citenamefont {Meyer}}]{Bonche1990npa}%
  \BibitemOpen
  \bibfield  {author} {\bibinfo {author} {\bibfnamefont {P.}~\bibnamefont
  {Bonche}}, \bibinfo {author} {\bibfnamefont {J.}~\bibnamefont {Dobaczewski}},
  \bibinfo {author} {\bibfnamefont {H.}~\bibnamefont {Flocard}}, \bibinfo
  {author} {\bibfnamefont {P.-H.}\ \bibnamefont {Heenen}}, \ and\ \bibinfo
  {author} {\bibfnamefont {J.}~\bibnamefont {Meyer}},\ }\href {\doibase
  https://doi.org/10.1016/0375-9474(90)90062-Q} {\bibfield  {journal} {\bibinfo
   {journal} {Nucl. Phys. A}\ }\textbf {\bibinfo {volume} {510}},\ \bibinfo
  {pages} {466} (\bibinfo {year} {1990})}\BibitemShut {NoStop}%
\bibitem [{\citenamefont {Bender}\ \emph {et~al.}(2006)\citenamefont {Bender},
  \citenamefont {Bonche},\ and\ \citenamefont {Heenen}}]{Bender2006prc}%
  \BibitemOpen
  \bibfield  {author} {\bibinfo {author} {\bibfnamefont {M.}~\bibnamefont
  {Bender}}, \bibinfo {author} {\bibfnamefont {P.}~\bibnamefont {Bonche}}, \
  and\ \bibinfo {author} {\bibfnamefont {P.-H.}\ \bibnamefont {Heenen}},\
  }\href {\doibase 10.1103/PhysRevC.74.024312} {\bibfield  {journal} {\bibinfo
  {journal} {Phys. Rev. C}\ }\textbf {\bibinfo {volume} {74}},\ \bibinfo
  {pages} {024312} (\bibinfo {year} {2006})}\BibitemShut {NoStop}%
\bibitem [{\citenamefont {Dobaczewski}\ \emph {et~al.}(2009)\citenamefont
  {Dobaczewski}, \citenamefont {Satula}, \citenamefont {Carlsson},
  \citenamefont {Engel}, \citenamefont {Olbratowski}, \citenamefont
  {Powalowski}, \citenamefont {Sadziak}, \citenamefont {Sarich}, \citenamefont
  {Schunck}, \citenamefont {Staszczak}, \citenamefont {Stoitsov}, \citenamefont
  {Zalewski},\ and\ \citenamefont {Zdunczuk}}]{Dobaczewski09}%
  \BibitemOpen
  \bibfield  {author} {\bibinfo {author} {\bibfnamefont {J.}~\bibnamefont
  {Dobaczewski}}, \bibinfo {author} {\bibfnamefont {W.}~\bibnamefont {Satula}},
  \bibinfo {author} {\bibfnamefont {B.}~\bibnamefont {Carlsson}}, \bibinfo
  {author} {\bibfnamefont {J.}~\bibnamefont {Engel}}, \bibinfo {author}
  {\bibfnamefont {P.}~\bibnamefont {Olbratowski}}, \bibinfo {author}
  {\bibfnamefont {P.}~\bibnamefont {Powalowski}}, \bibinfo {author}
  {\bibfnamefont {M.}~\bibnamefont {Sadziak}}, \bibinfo {author} {\bibfnamefont
  {J.}~\bibnamefont {Sarich}}, \bibinfo {author} {\bibfnamefont
  {N.}~\bibnamefont {Schunck}}, \bibinfo {author} {\bibfnamefont
  {A.}~\bibnamefont {Staszczak}}, \bibinfo {author} {\bibfnamefont
  {M.}~\bibnamefont {Stoitsov}}, \bibinfo {author} {\bibfnamefont
  {M.}~\bibnamefont {Zalewski}}, \ and\ \bibinfo {author} {\bibfnamefont
  {H.}~\bibnamefont {Zdunczuk}},\ }\href {\doibase
  https://doi.org/10.1016/j.cpc.2009.08.009} {\bibfield  {journal} {\bibinfo
  {journal} {Comput. Phys. Commun.}\ }\textbf {\bibinfo {volume} {180}},\
  \bibinfo {pages} {2361} (\bibinfo {year} {2009})}\BibitemShut {NoStop}%
\bibitem [{\citenamefont {Schunck}\ \emph {et~al.}(2012)\citenamefont
  {Schunck}, \citenamefont {Dobaczewski}, \citenamefont {McDonnell},
  \citenamefont {Satu{\l}a}, \citenamefont {Sheikh}, \citenamefont {Staszczak},
  \citenamefont {Stoitsov},\ and\ \citenamefont {Toivanen}}]{schunck2012CPC}%
  \BibitemOpen
  \bibfield  {author} {\bibinfo {author} {\bibfnamefont {N.}~\bibnamefont
  {Schunck}}, \bibinfo {author} {\bibfnamefont {J.}~\bibnamefont
  {Dobaczewski}}, \bibinfo {author} {\bibfnamefont {J.}~\bibnamefont
  {McDonnell}}, \bibinfo {author} {\bibfnamefont {W.}~\bibnamefont
  {Satu{\l}a}}, \bibinfo {author} {\bibfnamefont {J.}~\bibnamefont {Sheikh}},
  \bibinfo {author} {\bibfnamefont {A.}~\bibnamefont {Staszczak}}, \bibinfo
  {author} {\bibfnamefont {M.}~\bibnamefont {Stoitsov}}, \ and\ \bibinfo
  {author} {\bibfnamefont {P.}~\bibnamefont {Toivanen}},\ }\href@noop {}
  {\bibfield  {journal} {\bibinfo  {journal} {Comput. Phys. Commun.}\ }\textbf
  {\bibinfo {volume} {183}},\ \bibinfo {pages} {166} (\bibinfo {year}
  {2012})}\BibitemShut {NoStop}%
\bibitem [{\citenamefont {Cui}\ \emph {et~al.}(2017)\citenamefont {Cui},
  \citenamefont {Zhou}, \citenamefont {Guo},\ and\ \citenamefont
  {Schulze}}]{Cui2017prc}%
  \BibitemOpen
  \bibfield  {author} {\bibinfo {author} {\bibfnamefont {J.-W.}\ \bibnamefont
  {Cui}}, \bibinfo {author} {\bibfnamefont {X.-R.}\ \bibnamefont {Zhou}},
  \bibinfo {author} {\bibfnamefont {L.-X.}\ \bibnamefont {Guo}}, \ and\
  \bibinfo {author} {\bibfnamefont {H.-J.}\ \bibnamefont {Schulze}},\ }\href
  {\doibase 10.1103/PhysRevC.95.024323} {\bibfield  {journal} {\bibinfo
  {journal} {Phys. Rev. C}\ }\textbf {\bibinfo {volume} {95}},\ \bibinfo
  {pages} {024323} (\bibinfo {year} {2017})}\BibitemShut {NoStop}%
\bibitem [{\citenamefont {Li}\ \emph {et~al.}(2018)\citenamefont {Li},
  \citenamefont {Cui},\ and\ \citenamefont {Zhou}}]{Li2018prc}%
  \BibitemOpen
  \bibfield  {author} {\bibinfo {author} {\bibfnamefont {W.-Y.}\ \bibnamefont
  {Li}}, \bibinfo {author} {\bibfnamefont {J.-W.}\ \bibnamefont {Cui}}, \ and\
  \bibinfo {author} {\bibfnamefont {X.-R.}\ \bibnamefont {Zhou}},\ }\href
  {\doibase 10.1103/PhysRevC.97.034302} {\bibfield  {journal} {\bibinfo
  {journal} {Phys. Rev. C}\ }\textbf {\bibinfo {volume} {97}},\ \bibinfo
  {pages} {034302} (\bibinfo {year} {2018})}\BibitemShut {NoStop}%
\bibitem [{\citenamefont {Xue}\ \emph {et~al.}(2024{\natexlab{a}})\citenamefont
  {Xue}, \citenamefont {Cui}, \citenamefont {Chen}, \citenamefont {Zhou},\ and\
  \citenamefont {Sagawa}}]{Xue2024prc2}%
  \BibitemOpen
  \bibfield  {author} {\bibinfo {author} {\bibfnamefont {H.-T.}\ \bibnamefont
  {Xue}}, \bibinfo {author} {\bibfnamefont {J.-W.}\ \bibnamefont {Cui}},
  \bibinfo {author} {\bibfnamefont {Q.~B.}\ \bibnamefont {Chen}}, \bibinfo
  {author} {\bibfnamefont {X.-R.}\ \bibnamefont {Zhou}}, \ and\ \bibinfo
  {author} {\bibfnamefont {H.}~\bibnamefont {Sagawa}},\ }\href {\doibase
  10.1103/PhysRevC.110.044310} {\bibfield  {journal} {\bibinfo  {journal}
  {Phys. Rev. C}\ }\textbf {\bibinfo {volume} {110}},\ \bibinfo {pages}
  {044310} (\bibinfo {year} {2024}{\natexlab{a}})}\BibitemShut {NoStop}%
\bibitem [{\citenamefont {Xue}\ \emph {et~al.}(2024{\natexlab{b}})\citenamefont
  {Xue}, \citenamefont {Chen}, \citenamefont {Cui}, \citenamefont {Chen},
  \citenamefont {Schulze},\ and\ \citenamefont {Zhou}}]{Xue2024prc}%
  \BibitemOpen
  \bibfield  {author} {\bibinfo {author} {\bibfnamefont {H.-T.}\ \bibnamefont
  {Xue}}, \bibinfo {author} {\bibfnamefont {Q.~B.}\ \bibnamefont {Chen}},
  \bibinfo {author} {\bibfnamefont {J.-W.}\ \bibnamefont {Cui}}, \bibinfo
  {author} {\bibfnamefont {C.-F.}\ \bibnamefont {Chen}}, \bibinfo {author}
  {\bibfnamefont {H.-J.}\ \bibnamefont {Schulze}}, \ and\ \bibinfo {author}
  {\bibfnamefont {X.-R.}\ \bibnamefont {Zhou}},\ }\href {\doibase
  10.1103/PhysRevC.109.024324} {\bibfield  {journal} {\bibinfo  {journal}
  {Phys. Rev. C}\ }\textbf {\bibinfo {volume} {109}},\ \bibinfo {pages}
  {024324} (\bibinfo {year} {2024}{\natexlab{b}})}\BibitemShut {NoStop}%
\bibitem [{\citenamefont {Chen}\ \emph {et~al.}(2022)\citenamefont {Chen},
  \citenamefont {Chen}, \citenamefont {Zhou}, \citenamefont {Cheng},
  \citenamefont {Cui},\ and\ \citenamefont {Schulze}}]{C.F.Chen_cpc_2022}%
  \BibitemOpen
  \bibfield  {author} {\bibinfo {author} {\bibfnamefont {C.~F.}\ \bibnamefont
  {Chen}}, \bibinfo {author} {\bibfnamefont {Q.~B.}\ \bibnamefont {Chen}},
  \bibinfo {author} {\bibfnamefont {X.-R.}\ \bibnamefont {Zhou}}, \bibinfo
  {author} {\bibfnamefont {Y.~Y.}\ \bibnamefont {Cheng}}, \bibinfo {author}
  {\bibfnamefont {J.-W.}\ \bibnamefont {Cui}}, \ and\ \bibinfo {author}
  {\bibfnamefont {H.-J.}\ \bibnamefont {Schulze}},\ }\href@noop {} {\bibfield
  {journal} {\bibinfo  {journal} {Chin. Phys. C}\ }\textbf {\bibinfo {volume}
  {46}},\ \bibinfo {pages} {064109} (\bibinfo {year} {2022})}\BibitemShut
  {NoStop}%
\bibitem [{\citenamefont {Guo}\ \emph {et~al.}(2022)\citenamefont {Guo},
  \citenamefont {Chen}, \citenamefont {Zhou}, \citenamefont {Chen},\ and\
  \citenamefont {Schulze}}]{J.Guo_prc_2022}%
  \BibitemOpen
  \bibfield  {author} {\bibinfo {author} {\bibfnamefont {J.}~\bibnamefont
  {Guo}}, \bibinfo {author} {\bibfnamefont {C.~F.}\ \bibnamefont {Chen}},
  \bibinfo {author} {\bibfnamefont {X.-R.}\ \bibnamefont {Zhou}}, \bibinfo
  {author} {\bibfnamefont {Q.~B.}\ \bibnamefont {Chen}}, \ and\ \bibinfo
  {author} {\bibfnamefont {H.-J.}\ \bibnamefont {Schulze}},\ }\href {\doibase
  10.1103/PhysRevC.105.034322} {\bibfield  {journal} {\bibinfo  {journal}
  {Phys. Rev. C}\ }\textbf {\bibinfo {volume} {105}},\ \bibinfo {pages}
  {034322} (\bibinfo {year} {2022})}\BibitemShut {NoStop}%
\bibitem [{\citenamefont {Liu}\ \emph {et~al.}(2023)\citenamefont {Liu},
  \citenamefont {Chen}, \citenamefont {Chen}, \citenamefont {Xue},
  \citenamefont {Schulze},\ and\ \citenamefont {Zhou}}]{Y.Liu2023prc}%
  \BibitemOpen
  \bibfield  {author} {\bibinfo {author} {\bibfnamefont {Y.-X.}\ \bibnamefont
  {Liu}}, \bibinfo {author} {\bibfnamefont {C.~F.}\ \bibnamefont {Chen}},
  \bibinfo {author} {\bibfnamefont {Q.~B.}\ \bibnamefont {Chen}}, \bibinfo
  {author} {\bibfnamefont {H.-T.}\ \bibnamefont {Xue}}, \bibinfo {author}
  {\bibfnamefont {H.-J.}\ \bibnamefont {Schulze}}, \ and\ \bibinfo {author}
  {\bibfnamefont {X.-R.}\ \bibnamefont {Zhou}},\ }\href {\doibase
  10.1103/PhysRevC.108.064312} {\bibfield  {journal} {\bibinfo  {journal}
  {Phys. Rev. C}\ }\textbf {\bibinfo {volume} {108}},\ \bibinfo {pages}
  {064312} (\bibinfo {year} {2023})}\BibitemShut {NoStop}%
\bibitem [{\citenamefont {Reinhard}\ \emph {et~al.}(2011)\citenamefont
  {Reinhard}, \citenamefont {Maruhn}, \citenamefont {Umar},\ and\ \citenamefont
  {Oberacker}}]{Reinhard2011prc}%
  \BibitemOpen
  \bibfield  {author} {\bibinfo {author} {\bibfnamefont {P.-G.}\ \bibnamefont
  {Reinhard}}, \bibinfo {author} {\bibfnamefont {J.~A.}\ \bibnamefont
  {Maruhn}}, \bibinfo {author} {\bibfnamefont {A.~S.}\ \bibnamefont {Umar}}, \
  and\ \bibinfo {author} {\bibfnamefont {V.~E.}\ \bibnamefont {Oberacker}},\
  }\href {\doibase 10.1103/PhysRevC.83.034312} {\bibfield  {journal} {\bibinfo
  {journal} {Phys. Rev. C}\ }\textbf {\bibinfo {volume} {83}},\ \bibinfo
  {pages} {034312} (\bibinfo {year} {2011})}\BibitemShut {NoStop}%
\bibitem [{\citenamefont {Zhang}\ \emph {et~al.}(2016)\citenamefont {Zhang},
  \citenamefont {Schuetrumpf},\ and\ \citenamefont
  {Nazarewicz}}]{C.L.Zhang2016prc}%
  \BibitemOpen
  \bibfield  {author} {\bibinfo {author} {\bibfnamefont {C.~L.}\ \bibnamefont
  {Zhang}}, \bibinfo {author} {\bibfnamefont {B.}~\bibnamefont {Schuetrumpf}},
  \ and\ \bibinfo {author} {\bibfnamefont {W.}~\bibnamefont {Nazarewicz}},\
  }\href {\doibase 10.1103/PhysRevC.94.064323} {\bibfield  {journal} {\bibinfo
  {journal} {Phys. Rev. C}\ }\textbf {\bibinfo {volume} {94}},\ \bibinfo
  {pages} {064323} (\bibinfo {year} {2016})}\BibitemShut {NoStop}%
\bibitem [{\citenamefont {Schuetrumpf}\ and\ \citenamefont
  {Nazarewicz}(2017)}]{Schuetrumpf2017prc}%
  \BibitemOpen
  \bibfield  {author} {\bibinfo {author} {\bibfnamefont {B.}~\bibnamefont
  {Schuetrumpf}}\ and\ \bibinfo {author} {\bibfnamefont {W.}~\bibnamefont
  {Nazarewicz}},\ }\href {\doibase 10.1103/PhysRevC.96.064608} {\bibfield
  {journal} {\bibinfo  {journal} {Phys. Rev. C}\ }\textbf {\bibinfo {volume}
  {96}},\ \bibinfo {pages} {064608} (\bibinfo {year} {2017})}\BibitemShut
  {NoStop}%
\bibitem [{\citenamefont {Mercier}\ \emph {et~al.}(2021)\citenamefont
  {Mercier}, \citenamefont {Bjel\ifmmode \check{c}\else
  \v{c}\fi{}i\ifmmode~\acute{c}\else \'{c}\fi{}}, \citenamefont {Nik\ifmmode
  \check{s}\else \v{s}\fi{}i\ifmmode~\acute{c}\else \'{c}\fi{}}, \citenamefont
  {Ebran}, \citenamefont {Khan},\ and\ \citenamefont
  {Vretenar}}]{Mercier2021prc}%
  \BibitemOpen
  \bibfield  {author} {\bibinfo {author} {\bibfnamefont {F.}~\bibnamefont
  {Mercier}}, \bibinfo {author} {\bibfnamefont {A.}~\bibnamefont {Bjel\ifmmode
  \check{c}\else \v{c}\fi{}i\ifmmode~\acute{c}\else \'{c}\fi{}}}, \bibinfo
  {author} {\bibfnamefont {T.}~\bibnamefont {Nik\ifmmode \check{s}\else
  \v{s}\fi{}i\ifmmode~\acute{c}\else \'{c}\fi{}}}, \bibinfo {author}
  {\bibfnamefont {J.-P.}\ \bibnamefont {Ebran}}, \bibinfo {author}
  {\bibfnamefont {E.}~\bibnamefont {Khan}}, \ and\ \bibinfo {author}
  {\bibfnamefont {D.}~\bibnamefont {Vretenar}},\ }\href {\doibase
  10.1103/PhysRevC.103.024303} {\bibfield  {journal} {\bibinfo  {journal}
  {Phys. Rev. C}\ }\textbf {\bibinfo {volume} {103}},\ \bibinfo {pages}
  {024303} (\bibinfo {year} {2021})}\BibitemShut {NoStop}%
\bibitem [{\citenamefont {Li}\ \emph {et~al.}(2024)\citenamefont {Li},
  \citenamefont {Chen}, \citenamefont {Zhou},\ and\ \citenamefont
  {Ren}}]{Li2024prc}%
  \BibitemOpen
  \bibfield  {author} {\bibinfo {author} {\bibfnamefont {X.}~\bibnamefont
  {Li}}, \bibinfo {author} {\bibfnamefont {C.~F.}\ \bibnamefont {Chen}},
  \bibinfo {author} {\bibfnamefont {X.-R.}\ \bibnamefont {Zhou}}, \ and\
  \bibinfo {author} {\bibfnamefont {Z.}~\bibnamefont {Ren}},\ }\href {\doibase
  10.1103/PhysRevC.109.064301} {\bibfield  {journal} {\bibinfo  {journal}
  {Phys. Rev. C}\ }\textbf {\bibinfo {volume} {109}},\ \bibinfo {pages}
  {064301} (\bibinfo {year} {2024})}\BibitemShut {NoStop}%
\bibitem [{\citenamefont {Sagawa}\ and\ \citenamefont
  {Col\`{o}}(2014)}]{Sagawa2014PPNP}%
  \BibitemOpen
  \bibfield  {author} {\bibinfo {author} {\bibfnamefont {H.}~\bibnamefont
  {Sagawa}}\ and\ \bibinfo {author} {\bibfnamefont {G.}~\bibnamefont
  {Col\`{o}}},\ }\href {\doibase https://doi.org/10.1016/j.ppnp.2014.01.006}
  {\bibfield  {journal} {\bibinfo  {journal} {Prog. Part. Nucl. Phys.}\
  }\textbf {\bibinfo {volume} {76}},\ \bibinfo {pages} {76} (\bibinfo {year}
  {2014})}\BibitemShut {NoStop}%
\bibitem [{\citenamefont {Godbey}\ \emph {et~al.}(2019)\citenamefont {Godbey},
  \citenamefont {Guo},\ and\ \citenamefont {Umar}}]{Godbey2019prc}%
  \BibitemOpen
  \bibfield  {author} {\bibinfo {author} {\bibfnamefont {K.}~\bibnamefont
  {Godbey}}, \bibinfo {author} {\bibfnamefont {L.}~\bibnamefont {Guo}}, \ and\
  \bibinfo {author} {\bibfnamefont {A.~S.}\ \bibnamefont {Umar}},\ }\href
  {\doibase 10.1103/PhysRevC.100.054612} {\bibfield  {journal} {\bibinfo
  {journal} {Phys. Rev. C}\ }\textbf {\bibinfo {volume} {100}},\ \bibinfo
  {pages} {054612} (\bibinfo {year} {2019})}\BibitemShut {NoStop}%
\bibitem [{\citenamefont {Col\'{o}}\ \emph {et~al.}(2007)\citenamefont
  {Col\'{o}}, \citenamefont {Sagawa}, \citenamefont {Fracasso},\ and\
  \citenamefont {Bortignon}}]{Colo2007PLB}%
  \BibitemOpen
  \bibfield  {author} {\bibinfo {author} {\bibfnamefont {G.}~\bibnamefont
  {Col\'{o}}}, \bibinfo {author} {\bibfnamefont {H.}~\bibnamefont {Sagawa}},
  \bibinfo {author} {\bibfnamefont {S.}~\bibnamefont {Fracasso}}, \ and\
  \bibinfo {author} {\bibfnamefont {P.}~\bibnamefont {Bortignon}},\ }\href
  {\doibase https://doi.org/10.1016/j.physletb.2007.01.033} {\bibfield
  {journal} {\bibinfo  {journal} {Phys. Lett. B}\ }\textbf {\bibinfo {volume}
  {646}},\ \bibinfo {pages} {227} (\bibinfo {year} {2007})}\BibitemShut
  {NoStop}%
\bibitem [{\citenamefont {Brink}\ and\ \citenamefont
  {Stancu}(2007)}]{Brink2007PRC}%
  \BibitemOpen
  \bibfield  {author} {\bibinfo {author} {\bibfnamefont {D.~M.}\ \bibnamefont
  {Brink}}\ and\ \bibinfo {author} {\bibfnamefont {F.}~\bibnamefont {Stancu}},\
  }\href {\doibase 10.1103/PhysRevC.75.064311} {\bibfield  {journal} {\bibinfo
  {journal} {Phys. Rev. C}\ }\textbf {\bibinfo {volume} {75}},\ \bibinfo
  {pages} {064311} (\bibinfo {year} {2007})}\BibitemShut {NoStop}%
\bibitem [{\citenamefont {Lesinski}\ \emph {et~al.}(2007)\citenamefont
  {Lesinski}, \citenamefont {Bender}, \citenamefont {Bennaceur}, \citenamefont
  {Duguet},\ and\ \citenamefont {Meyer}}]{TIJ}%
  \BibitemOpen
  \bibfield  {author} {\bibinfo {author} {\bibfnamefont {T.}~\bibnamefont
  {Lesinski}}, \bibinfo {author} {\bibfnamefont {M.}~\bibnamefont {Bender}},
  \bibinfo {author} {\bibfnamefont {K.}~\bibnamefont {Bennaceur}}, \bibinfo
  {author} {\bibfnamefont {T.}~\bibnamefont {Duguet}}, \ and\ \bibinfo {author}
  {\bibfnamefont {J.}~\bibnamefont {Meyer}},\ }\href {\doibase
  10.1103/PhysRevC.76.014312} {\bibfield  {journal} {\bibinfo  {journal} {Phys.
  Rev. C}\ }\textbf {\bibinfo {volume} {76}},\ \bibinfo {pages} {014312}
  (\bibinfo {year} {2007})}\BibitemShut {NoStop}%
\bibitem [{\citenamefont {Shen}\ \emph {et~al.}(2019)\citenamefont {Shen},
  \citenamefont {Col\`o},\ and\ \citenamefont {Roca-Maza}}]{SAMiT}%
  \BibitemOpen
  \bibfield  {author} {\bibinfo {author} {\bibfnamefont {S.}~\bibnamefont
  {Shen}}, \bibinfo {author} {\bibfnamefont {G.}~\bibnamefont {Col\`o}}, \ and\
  \bibinfo {author} {\bibfnamefont {X.}~\bibnamefont {Roca-Maza}},\ }\href
  {\doibase 10.1103/PhysRevC.99.034322} {\bibfield  {journal} {\bibinfo
  {journal} {Phys. Rev. C}\ }\textbf {\bibinfo {volume} {99}},\ \bibinfo
  {pages} {034322} (\bibinfo {year} {2019})}\BibitemShut {NoStop}%
\bibitem [{\citenamefont {Ring}\ and\ \citenamefont
  {Schuck}(1980)}]{Peter1980}%
  \BibitemOpen
  \bibfield  {author} {\bibinfo {author} {\bibfnamefont {P.}~\bibnamefont
  {Ring}}\ and\ \bibinfo {author} {\bibfnamefont {P.}~\bibnamefont {Schuck}},\
  }\href@noop {} {\emph {\bibinfo {title} {The nuclear many body problem}}}\
  (\bibinfo  {publisher} {Springer Verlag, Berlin},\ \bibinfo {year}
  {1980})\BibitemShut {NoStop}%
\bibitem [{\citenamefont {Dobaczewski}\ \emph {et~al.}(1984)\citenamefont
  {Dobaczewski}, \citenamefont {Flocard},\ and\ \citenamefont {Treiner}}]{skp}%
  \BibitemOpen
  \bibfield  {author} {\bibinfo {author} {\bibfnamefont {J.}~\bibnamefont
  {Dobaczewski}}, \bibinfo {author} {\bibfnamefont {H.}~\bibnamefont
  {Flocard}}, \ and\ \bibinfo {author} {\bibfnamefont {J.}~\bibnamefont
  {Treiner}},\ }\href {\doibase https://doi.org/10.1016/0375-9474(84)90433-0}
  {\bibfield  {journal} {\bibinfo  {journal} {Nucl. Phys. A}\ }\textbf
  {\bibinfo {volume} {422}},\ \bibinfo {pages} {103} (\bibinfo {year}
  {1984})}\BibitemShut {NoStop}%
\bibitem [{\citenamefont {Chabanat}\ \emph {et~al.}(1998)\citenamefont
  {Chabanat}, \citenamefont {Bonche}, \citenamefont {Haensel}, \citenamefont
  {Meyer},\ and\ \citenamefont {Schaeffer}}]{Chabanat1998npa}%
  \BibitemOpen
  \bibfield  {author} {\bibinfo {author} {\bibfnamefont {E.}~\bibnamefont
  {Chabanat}}, \bibinfo {author} {\bibfnamefont {P.}~\bibnamefont {Bonche}},
  \bibinfo {author} {\bibfnamefont {P.}~\bibnamefont {Haensel}}, \bibinfo
  {author} {\bibfnamefont {J.}~\bibnamefont {Meyer}}, \ and\ \bibinfo {author}
  {\bibfnamefont {R.}~\bibnamefont {Schaeffer}},\ }\href {\doibase
  https://doi.org/10.1016/S0375-9474(98)00180-8} {\bibfield  {journal}
  {\bibinfo  {journal} {Nucl. Phys. A}\ }\textbf {\bibinfo {volume} {635}},\
  \bibinfo {pages} {231} (\bibinfo {year} {1998})}\BibitemShut {NoStop}%
\bibitem [{\citenamefont {Reinhard}\ \emph {et~al.}(1999)\citenamefont
  {Reinhard}, \citenamefont {Dean}, \citenamefont {Nazarewicz}, \citenamefont
  {Dobaczewski}, \citenamefont {Maruhn},\ and\ \citenamefont {Strayer}}]{sko'}%
  \BibitemOpen
  \bibfield  {author} {\bibinfo {author} {\bibfnamefont {P.-G.}\ \bibnamefont
  {Reinhard}}, \bibinfo {author} {\bibfnamefont {D.~J.}\ \bibnamefont {Dean}},
  \bibinfo {author} {\bibfnamefont {W.}~\bibnamefont {Nazarewicz}}, \bibinfo
  {author} {\bibfnamefont {J.}~\bibnamefont {Dobaczewski}}, \bibinfo {author}
  {\bibfnamefont {J.~A.}\ \bibnamefont {Maruhn}}, \ and\ \bibinfo {author}
  {\bibfnamefont {M.~R.}\ \bibnamefont {Strayer}},\ }\href {\doibase
  10.1103/PhysRevC.60.014316} {\bibfield  {journal} {\bibinfo  {journal} {Phys.
  Rev. C}\ }\textbf {\bibinfo {volume} {60}},\ \bibinfo {pages} {014316}
  (\bibinfo {year} {1999})}\BibitemShut {NoStop}%
\bibitem [{\citenamefont {Goriely}\ \emph {et~al.}(2005)\citenamefont
  {Goriely}, \citenamefont {Samyn}, \citenamefont {Pearson},\ and\
  \citenamefont {Onsi}}]{BSk9}%
  \BibitemOpen
  \bibfield  {author} {\bibinfo {author} {\bibfnamefont {S.}~\bibnamefont
  {Goriely}}, \bibinfo {author} {\bibfnamefont {M.}~\bibnamefont {Samyn}},
  \bibinfo {author} {\bibfnamefont {J.}~\bibnamefont {Pearson}}, \ and\
  \bibinfo {author} {\bibfnamefont {M.}~\bibnamefont {Onsi}},\ }\href {\doibase
  https://doi.org/10.1016/j.nuclphysa.2005.01.009} {\bibfield  {journal}
  {\bibinfo  {journal} {Nuclear Physics A}\ }\textbf {\bibinfo {volume}
  {750}},\ \bibinfo {pages} {425} (\bibinfo {year} {2005})}\BibitemShut
  {NoStop}%
\bibitem [{\citenamefont {Roca-Maza}\ \emph {et~al.}(2012)\citenamefont
  {Roca-Maza}, \citenamefont {Col\`o},\ and\ \citenamefont {Sagawa}}]{SAMi}%
  \BibitemOpen
  \bibfield  {author} {\bibinfo {author} {\bibfnamefont {X.}~\bibnamefont
  {Roca-Maza}}, \bibinfo {author} {\bibfnamefont {G.}~\bibnamefont {Col\`o}}, \
  and\ \bibinfo {author} {\bibfnamefont {H.}~\bibnamefont {Sagawa}},\ }\href
  {\doibase 10.1103/PhysRevC.86.031306} {\bibfield  {journal} {\bibinfo
  {journal} {Phys. Rev. C}\ }\textbf {\bibinfo {volume} {86}},\ \bibinfo
  {pages} {031306} (\bibinfo {year} {2012})}\BibitemShut {NoStop}%
\bibitem [{\citenamefont {Brown}\ \emph {et~al.}(2006)\citenamefont {Brown},
  \citenamefont {Duguet}, \citenamefont {Otsuka}, \citenamefont {Abe},\ and\
  \citenamefont {Suzuki}}]{Skxtb}%
  \BibitemOpen
  \bibfield  {author} {\bibinfo {author} {\bibfnamefont {B.~A.}\ \bibnamefont
  {Brown}}, \bibinfo {author} {\bibfnamefont {T.}~\bibnamefont {Duguet}},
  \bibinfo {author} {\bibfnamefont {T.}~\bibnamefont {Otsuka}}, \bibinfo
  {author} {\bibfnamefont {D.}~\bibnamefont {Abe}}, \ and\ \bibinfo {author}
  {\bibfnamefont {T.}~\bibnamefont {Suzuki}},\ }\href {\doibase
  10.1103/PhysRevC.74.061303} {\bibfield  {journal} {\bibinfo  {journal} {Phys.
  Rev. C}\ }\textbf {\bibinfo {volume} {74}},\ \bibinfo {pages} {061303}
  (\bibinfo {year} {2006})}\BibitemShut {NoStop}%
\bibitem [{\citenamefont {Zalewski}\ \emph {et~al.}(2008)\citenamefont
  {Zalewski}, \citenamefont {Dobaczewski}, \citenamefont {Satu\l{}a},\ and\
  \citenamefont {Werner}}]{SLy4T}%
  \BibitemOpen
  \bibfield  {author} {\bibinfo {author} {\bibfnamefont {M.}~\bibnamefont
  {Zalewski}}, \bibinfo {author} {\bibfnamefont {J.}~\bibnamefont
  {Dobaczewski}}, \bibinfo {author} {\bibfnamefont {W.}~\bibnamefont
  {Satu\l{}a}}, \ and\ \bibinfo {author} {\bibfnamefont {T.~R.}\ \bibnamefont
  {Werner}},\ }\href {\doibase 10.1103/PhysRevC.77.024316} {\bibfield
  {journal} {\bibinfo  {journal} {Phys. Rev. C}\ }\textbf {\bibinfo {volume}
  {77}},\ \bibinfo {pages} {024316} (\bibinfo {year} {2008})}\BibitemShut
  {NoStop}%
\bibitem [{\citenamefont {Beiner}\ \emph {et~al.}(1975)\citenamefont {Beiner},
  \citenamefont {Flocard}, \citenamefont {{Van Giai}},\ and\ \citenamefont
  {Quentin}}]{Beiner1975npa}%
  \BibitemOpen
  \bibfield  {author} {\bibinfo {author} {\bibfnamefont {M.}~\bibnamefont
  {Beiner}}, \bibinfo {author} {\bibfnamefont {H.}~\bibnamefont {Flocard}},
  \bibinfo {author} {\bibfnamefont {N.}~\bibnamefont {{Van Giai}}}, \ and\
  \bibinfo {author} {\bibfnamefont {P.}~\bibnamefont {Quentin}},\ }\href
  {\doibase https://doi.org/10.1016/0375-9474(75)90338-3} {\bibfield  {journal}
  {\bibinfo  {journal} {Nucl. Phys. A}\ }\textbf {\bibinfo {volume} {238}},\
  \bibinfo {pages} {29} (\bibinfo {year} {1975})}\BibitemShut {NoStop}%
\bibitem [{\citenamefont {Kortelainen}\ \emph {et~al.}(2010)\citenamefont
  {Kortelainen}, \citenamefont {Lesinski}, \citenamefont {Mor\'e},
  \citenamefont {Nazarewicz}, \citenamefont {Sarich}, \citenamefont {Schunck},
  \citenamefont {Stoitsov},\ and\ \citenamefont {Wild}}]{UDF0}%
  \BibitemOpen
  \bibfield  {author} {\bibinfo {author} {\bibfnamefont {M.}~\bibnamefont
  {Kortelainen}}, \bibinfo {author} {\bibfnamefont {T.}~\bibnamefont
  {Lesinski}}, \bibinfo {author} {\bibfnamefont {J.}~\bibnamefont {Mor\'e}},
  \bibinfo {author} {\bibfnamefont {W.}~\bibnamefont {Nazarewicz}}, \bibinfo
  {author} {\bibfnamefont {J.}~\bibnamefont {Sarich}}, \bibinfo {author}
  {\bibfnamefont {N.}~\bibnamefont {Schunck}}, \bibinfo {author} {\bibfnamefont
  {M.~V.}\ \bibnamefont {Stoitsov}}, \ and\ \bibinfo {author} {\bibfnamefont
  {S.}~\bibnamefont {Wild}},\ }\href {\doibase 10.1103/PhysRevC.82.024313}
  {\bibfield  {journal} {\bibinfo  {journal} {Phys. Rev. C}\ }\textbf {\bibinfo
  {volume} {82}},\ \bibinfo {pages} {024313} (\bibinfo {year}
  {2010})}\BibitemShut {NoStop}%
\bibitem [{\citenamefont {{Van Giai}}\ and\ \citenamefont
  {Sagawa}(1981)}]{Vangiai1981npa}%
  \BibitemOpen
  \bibfield  {author} {\bibinfo {author} {\bibfnamefont {N.}~\bibnamefont {{Van
  Giai}}}\ and\ \bibinfo {author} {\bibfnamefont {H.}~\bibnamefont {Sagawa}},\
  }\href {\doibase https://doi.org/10.1016/0375-9474(81)90741-7} {\bibfield
  {journal} {\bibinfo  {journal} {Nucl. Phys. A}\ }\textbf {\bibinfo {volume}
  {371}},\ \bibinfo {pages} {1} (\bibinfo {year} {1981})}\BibitemShut {NoStop}%
\bibitem [{\citenamefont {Sagawa}\ \emph {et~al.}(2004)\citenamefont {Sagawa},
  \citenamefont {Zhou}, \citenamefont {Zhang},\ and\ \citenamefont
  {Suzuki}}]{Sagawa2004PRC}%
  \BibitemOpen
  \bibfield  {author} {\bibinfo {author} {\bibfnamefont {H.}~\bibnamefont
  {Sagawa}}, \bibinfo {author} {\bibfnamefont {X.~R.}\ \bibnamefont {Zhou}},
  \bibinfo {author} {\bibfnamefont {X.~Z.}\ \bibnamefont {Zhang}}, \ and\
  \bibinfo {author} {\bibfnamefont {T.}~\bibnamefont {Suzuki}},\ }\href
  {\doibase 10.1103/PhysRevC.70.054316} {\bibfield  {journal} {\bibinfo
  {journal} {Phys. Rev. C}\ }\textbf {\bibinfo {volume} {70}},\ \bibinfo
  {pages} {054316} (\bibinfo {year} {2004})}\BibitemShut {NoStop}%
\bibitem [{\citenamefont {{Terasaki}}\ \emph {et~al.}(1996)\citenamefont
  {{Terasaki}}, \citenamefont {{Heenen}}, \citenamefont {{Flocard}},\ and\
  \citenamefont {{Bonche}}}]{Terasaki96}%
  \BibitemOpen
  \bibfield  {author} {\bibinfo {author} {\bibfnamefont {J.}~\bibnamefont
  {{Terasaki}}}, \bibinfo {author} {\bibfnamefont {P.~H.}\ \bibnamefont
  {{Heenen}}}, \bibinfo {author} {\bibfnamefont {H.}~\bibnamefont {{Flocard}}},
  \ and\ \bibinfo {author} {\bibfnamefont {P.}~\bibnamefont {{Bonche}}},\
  }\href {\doibase 10.1016/0375-9474(96)00036-X} {\bibfield  {journal}
  {\bibinfo  {journal} {\nphysa}\ }\textbf {\bibinfo {volume} {600}},\ \bibinfo
  {pages} {371} (\bibinfo {year} {1996})}\BibitemShut {NoStop}%
\end{thebibliography}%

\end{CJK*}

\end{document}